\documentclass[fleqn,usenatbib]{mnras}

\usepackage{newtxtext,newtxmath}
\usepackage[T1]{fontenc}
\usepackage{ae,aecompl}

\usepackage{graphicx}	% Including figure files
\usepackage{amsmath}	% Advanced maths commands
\usepackage{amssymb}	% Extra maths symbols
\usepackage{tikz}
\usepackage[utf8]{inputenc}

\newcommand{\msun}{~\mathrm{M_\odot}}
\newcommand{\au}{~\mathrm{au}}
\newcommand{\pc}{~\mathrm{pc}}
\title[Indirect evidence of significant grain growth in young protostellar envelopes]{Indirect evidence of significant grain growth in young protostellar envelopes from polarized dust emission}

\author[V. Valdivia et al.]{
Valeska Valdivia,$^{1}$\thanks{E-mail: valeska.valdivia@cea.fr}
Ana\"elle Maury,$^{1}$
Robert Brauer,$^{1}$
Patrick Hennebelle,$^{1}$
\newauthor
Maud Galametz,$^{1}$
Vincent Guillet,$^{2, 3}$
and Stefan Reissl$^{4}$
\\
$^{1}$Laboratoire AIM, Paris-Saclay, CEA/IRFU/SAp - CNRS - Universit\'e Paris Diderot, 91191 Gif-sur-Yvette Cedex, France\\
$^{2}$Institut d'Astrophysique Spatiale, CNRS, Univ. Paris-Sud, Univ. Paris-Saclay, B\^at. 121, 91405 Orsay Cedex, France\\
$^{3}${LUMP, Universit\'e de Montpellier, CNRS/IN2P3, CC 72, Place Eug\`ene Bataillon,
34095 Montpellier Cedex 5, France}\\
$^{4}$Universit\"at Heidelberg, Zentrum f\"ur Astronomie, ITA, Albert-Ueberle-Str. 2, 69120 Heidelberg, Germany
}

\date{Accepted XXX. Received YYY; in original form ZZZ}

\pubyear{2019}

\begin{document}
\label{firstpage}
\pagerange{\pageref{firstpage}--\pageref{lastpage}}
\maketitle

\begin{abstract}
How and when in the star formation sequence do dust grains start to grow into pebbles is a cornerstone question to both star and planet formation. 
We compute the polarized radiative transfer from a model solar-type protostellar core, using the \textsc{POLARIS} code, aligning the dust grains with the local magnetic field, following the radiative torques (RATs) theory. We test the dependency of the resulting dust polarized emission with the maximum grain size of the dust size distribution at the envelope scale, from $a_\mathrm{max}=1~\micron$ to $50 ~\micron$. 
Our work shows that, in the framework of RAT alignment, large dust grains are required to produce polarized dust emission at levels similar to those currently observed in solar-type protostellar envelopes at millimeter wavelengths. Considering the current theoretical difficulties to align a large fraction of small ISM-like grains in the conditions typical of protostellar envelopes, our results suggest that grain growth (typically $>10 ~\micron$) might have already significantly progressed at scales $100-1000 \au$ in the youngest objects, observed less than $10^5$ years after the onset of collapse. Observations of dust polarized emission might open a new avenue to explore dust pristine properties and describe, for example, the initial conditions for the formation of planetesimals.
\end{abstract}


%%%%%%%%%%%%%%%%%%%%%%%%%%%%%%%%%%%%%%%%%%%%%%%%%%

%%%%%%%%%%%%%%%%% BODY OF PAPER %%%%%%%%%%%%%%%%%%

\section{Introduction}
%\subsection{Current constraints on grain growth along the sequence leading to planet-bearing disks}

%In our quest to understand both star and planet formation, one cornerstone question is how dust grains evolve from their pristine properties in the interstellar medium (ISM) reservoir to planetesimals in Class II disks \citep{Testi2014}. Towards protoplanetary disks, the observed dust spectral distributions are much shallower than in the diffuse ISM \citep{Isella2010, Perez2015, Tripathi2017}, with an emissivity spectral index decreasing from the outer to the inner disk regions \citep{Guilloteau2011, Tazzari2016}. These observational clues suggest that grains from subnanometric sizes to cm pebbles co-exist in protoplanetary disks \citep{Natta2004, Birnstiel2010P}.
In our quest to understand both star and planet formation, one cornerstone question is how dust grains evolve from their pristine properties in the interstellar medium (ISM) to planetesimals in Class II disks, where observations suggest that grains from subnanometric sizes to cm pebbles co-exist \citep[see, e.g.][for a review]{Testi2014}.
While understanding the transition from pebbles to planets is the focus of many circumstellar disk studies, characterizing the early stages of grain growth is at least as important, not only to put constraints on the initial dust grain size distribution for disk evolution \citep[see, e.g.][for a review]{Birnstiel2016}, but to address several key issues such as the formation of complex organic molecules \citep[see, e.g.][for a review]{Herbst2017} or characterize the coupling of the magnetic field with the gas in star-forming cores \citep[e.g.][]{Zhao2018}.
A remaining open question is how and when the initial fractal growth of ISM grains to micronic aggregates proceeds, from which grains can further grow by grain-grain collisions and ram pressure of the gas \citep{Blum2008, Dominik2016}. Analytical and numerical models of grain growth in cores \citep{Ormel2011,ChaconTanarro2017} suggested that ISM grains can grow up to $\sim 100~\micron$ sizes over the estimated lifetime of prestellar cores \citep[1 Myr, e.g.][]{Konyves2015}.
In molecular clouds and dense cold cores, the observed decrease in the mid-IR to far-IR emission ratio \citep{Stepnik2003,Flagey2009}, the increase in the dust far-IR/millimeter opacity and spectral index \citep{Martin2012,Bracco2017}, and the increased scattering efficiency from the near- to mid-IR \citep[][]{Pagani2010} point out to an evolution of dust properties in dense regions. It was suggested that these observations could be explained if dense structures contain aliphatic-rich ice-mantled grains \citep{Jones2013,Ysard2016}, or already evolved micrometer-sized grains \citep{Steinacker2010,Lefevre2016}.\\
A rather unexplored step on the evolutionary sequence leading to the formation of planets is the protostellar stage, and even more so the main accretion phase \citep[Class\,0 protostars, ][]{Andre2000}. 
During this short ($<10^5$ years) but cornerstone phase for both the formation of stars and disks, half of the final stellar mass is accreted from the surrounding envelope into the stellar embryo: in about a free-fall time a typical dust grain (coupled to the gas) will drift from the outer protostellar envelope towards the much higher density, higher temperature, inner region. 
While the early stages of the formation and evolution of disks around protostars remain a very active research topic both observationally and theoretically \citep{Maury2019, Wurster2018}, it is yet very difficult to find robust tracers of the dust properties in these Class\,0 disks and envelopes: only a few studies suggested dust spectral indices compatible with grain growth \citep[in 3 objects,][]{Kwon2009}. 
In the presence of a magnetic field, non-spherical spinning dust grains will precess around magnetic field lines \citep{Davis51,Purcell79}, aligning their long axis perpendicular to the local magnetic field lines, producing a polarized thermal dust emission \citep[][]{Roberge2004}. Among the mechanisms able to align the dust grains, the Radiative Alignment Torques \citep[RATs,][]{Draine97, Hoang14}, has become recently the favoured alignment process in the diffuse ISM \cite[see][and references therein]{Andersson2015}: in the current RATs theory, the absorption and scattering of photons exert a differential torque that spins up, precesses and aligns helical dust grains. 
The RATs alignment torques act on short timescales ($10^3-10^4~\mathrm{yr}$) and are more efficient than the magnetic relaxation \citep{Draine97}. 
The efficiency of the RATs mechanism depends, among other parameters, on the strength and spectral energy distribution of the incoming radiation field, and thus on the size of the dust grains \citep{Draine1996, Cho2007}: in this paper we study the potential of the polarized dust emission, in the RATs framework, to constrain the dust size distribution in protostellar cores. 

%wavelength and the angle between the incident radiation and the magnetic field, on the dust size, on the wavelength of the photons c
%\begin{comment}
%- RAT depends on incident radiation vs B orientation, on T. 
%- Rat mechanism is sensitive to the dust size since bigger grains are better at aligning\\
%- Can the polarized emission help us to constrain the dust size distribution?\\
%- Realistic density distribution from a simulation, not a model...
%\end{comment}

%In this paper we study the effect of the dust size distribution on the polarized dust thermal emission.
%RAT

%Dust distribution 

%__________________________________________________________________

%RAT, dust pop and assumptions 
\section{Polarized dust radiative transfer} % from MHD simulations with \textsc{POLARIS}}
\label{res_synth_pol}

%To test the effect of the dust size distribution on the resulting polarized dust emission produced when aligning grains with the RATs mechanism, we use a non-ideal magnetohydrodynamical (MHD) simulation performed with the \textsc{RAMSES} code \citep{Teyssier2002, Fromang2006}. The simulation is described in more detail in \citet{Maury2018}: it follows the magnetized collapse of a $2.5 \msun$ dense core  with a mass-to-flux ratio $\mu=6$ and a ratio between the rotational and  gravitational energy of $E_\mathrm{rot}/ E_\mathrm{grav}=10^{-3}$. The full size of the simulation is $0.167 \pc$ ($\sim 34481.6 \au$) and the effective resolution is roughly $1 \au$.\\
To test the effect of the dust size distribution we post-process a snapshot of a non-ideal magnetohydrodynamical (MHD) simulation performed with the \textsc{RAMSES} code \citep{Teyssier2002, Fromang2006}. The simulation, described in more detail in Valdivia et al. (in prep), follows the magnetized collapse of a $2.5 \msun$ dense core  with a mass-to-flux ratio $\mu=6.6$ and a ratio between the rotational and  gravitational energy of $E_\mathrm{rot}/ E_\mathrm{grav}=10^{-2}$. The full size of the simulation is $0.167 \pc$ ($\sim 34481.6 \au$) and the effective resolution is $\sim 1 \au$.\\
We selected a time step where the protostar has already accreted $0.2 \msun$, which reproduces the typical conditions of a young protostar in its main accretion phase. Figure~\ref{input} shows the total column density ($N$), as well as the integrated magnetic field strength ($\textit{\textbf{B}}$) and orientation. Additionally, we provide the mean radial profile of the gas density in Fig.~\ref{profile_dens}.
%
%
\iffalse
\begin{figure}
\centering
\includegraphics[width=0.45\textwidth, trim={4.63cm 3.0cm 0.1cm 1.2cm},clip]{Figures/NBg_y_full10242_00051.pdf}
\caption{Column density distribution (background gray scale), from the modeled protostellar core at a scale of $2000 \au$ (the core extends over 7000 au in our simulation). Note that the core is seen almost edge-on, with the outflow lying in the plane of the sky, creating the north-south bubbles seen here). Overlaid are the mean magnetic field lines orientation, color-coded by magnetic field intensity (both obtained from line-of-sight integration, local mass-weighted, of the magnetic field projected onto the plane of the sky).}
\label{input}
\end{figure}
\fi
%
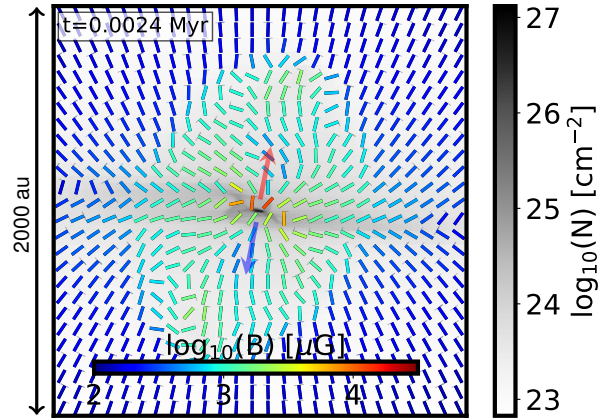
\begin{figure}
\centering
%\hspace{0.4cm}
\begin{tikzpicture}
\node[above right] (img) at (0,0) {\includegraphics[width=0.48\textwidth, trim={4.cm 3.0cm 0.1cm 1.8cm},clip]{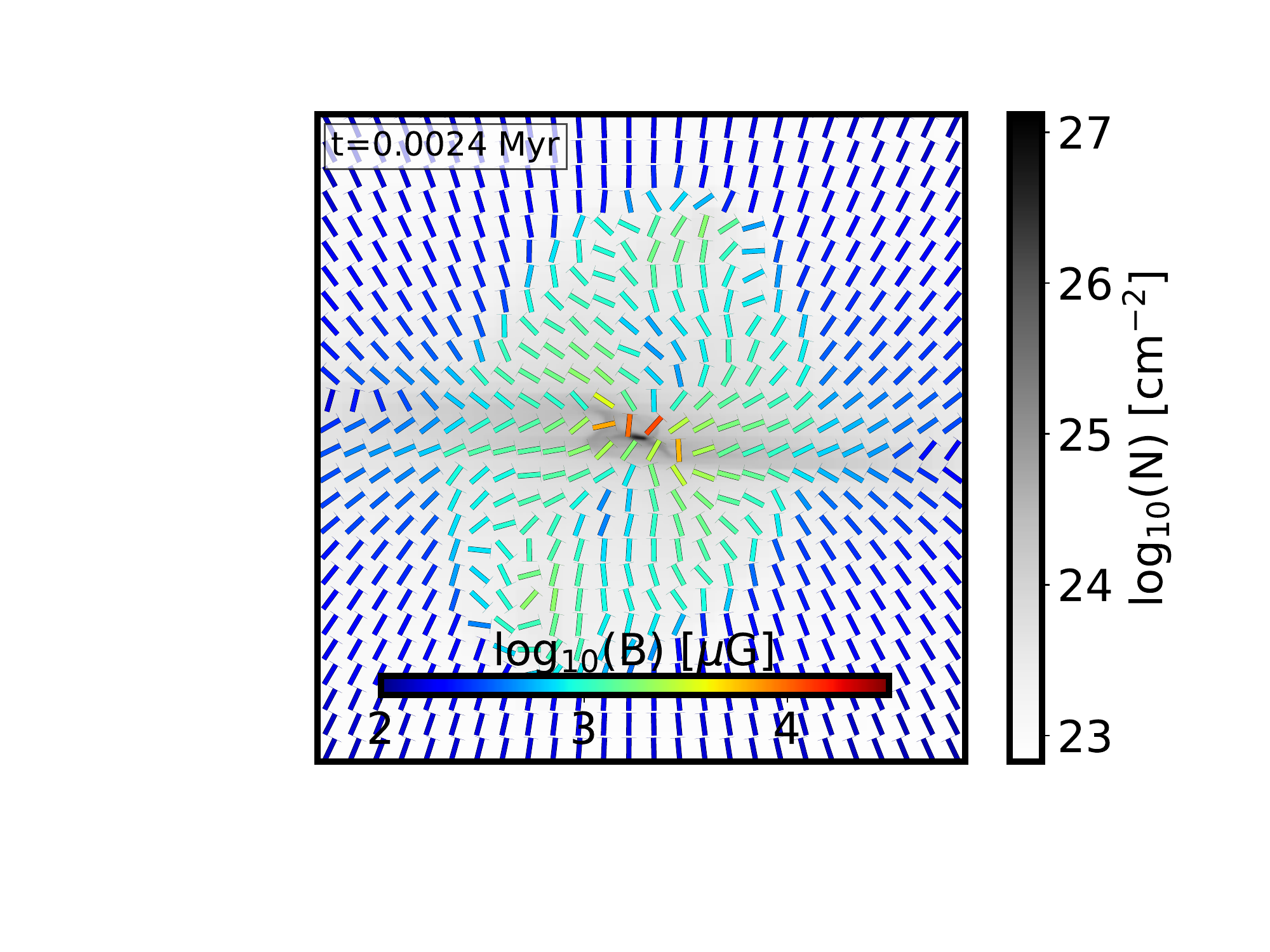}};
%\node at (0pt,0pt) {$-1000~\au$};
%\draw [<->, line width=1.2pt] (0.35,0.145) -- (0.35,5.3);
%\node [rotate=90] at (0.15,2.8) {\fontfamily{phv}\selectfont 2000 au};
\draw [<->, line width=1.2pt] (0.45,0.145) -- (0.45,5.55);
\node [rotate=90] at (0.25,2.8) {\fontfamily{phv}\selectfont 2000 au};
%\node [double arrow,draw, text width=5cm,align=center];
%\draw  (0,0) -- (1,0) edge [double] ++(10pt,0);
%\draw [->, double, line width=1.2pt, double distance=2pt] (2.78,2.48) -- (3.1,3.4);
%\draw [->, double, line width=1.2pt, double distance=2pt] (2.75,2.38) -- (2.43,1.48);
\begin{scope}[transparency group, opacity=0.5]
\draw[-stealth,line width=1.8pt, red] (3.39,3.0) -- (3.53,3.7);
\draw[-stealth,line width=1.8pt, blue] (3.34,2.7) -- (3.2,2);
\end{scope}
\end{tikzpicture}
\vspace{-0.3cm}
\caption{Column density distribution (background gray scale), from the modeled protostellar core at a scale of $2000 \au$ (the core extends over 7000 au in our simulation). Note that the core is seen almost edge-on, with the outflow lying in the plane of the sky, creating the north-south bubbles seen here. Overlaid are the mean magnetic field lines orientation, color-coded by magnetic field intensity (both obtained from line-of-sight integration, local mass-weighted, of the magnetic field projected onto the plane of the sky). The arrows indicate the outflows direction.} % as in \citet{Maury2018}
	\label{input}
\end{figure}

%%%%%%%%%%%%%%%%%%%%%%%%%%%%

%We compute the radiative transfer from the \textsc{RAMSES} output, using the \textsc{POLARIS} code \citep{Reissl17}, for the dust thermal emission and its polarization.
We perform the radiative transfer using the \textsc{POLARIS} code \citep{Reissl16}, capable of dust heating, thermal emission, and calculating the grain alignment efficiency of non-spherical dust grains with the local magnetic field lines following the RATs theory of \citet{LazarianHoang07} as outlined in \citet{Hoang14}.
We model the central source as a blackbody of $2~\mathrm{L_\odot}$ (of radius $1~\mathrm{R_\odot}$), similar to typical accretion luminosities observed towards low-mass young protostars \citep{Dunham2014}. 
We also include the interstellar radiation field (ISRF) as an external radiation field of strength $G_0=1$, using the \citet{Mathis1983} description. %\\
We assume a gas-to-dust mass ratio of $100$, and a composition of $62.5\%$ astronomical silicates and $37.5\%$ graphite grains, that fits the Galactic extinction curve \citep{Mathis1977}. The dust grains are supposed oblate, with an aspect ratio of $0.5$ \citep{Hildebrand95}. We use a MRN-like dust size distribution \citep{Mathis1977}, described by $\mathrm{d}n(a)\propto a^{-3.5}\mathrm{d}a$, where $a$, the effective radius of the grain, is the radius of a sphere of equivalent volume, and $n(a)$ is the number of dust grains of effective radius $a$. 
To test the influence of the presence of larger grains on the polarized dust emission we set the minimum dust grain size to $a_\mathrm{min} = 5~\mathrm{nm}$, and we vary the maximum size of the distribution from $a_\mathrm{max}=1~\micron$ to $50 ~\micron$ keeping the total dust mass constant. Our simulation reproduces the standard opacities observed in embedded protostars, resulting in a radiation field dominated by low energy photons at the envelope scales. The size distribution and the choice of the dust composition used here do not greatly impact the opacity.
We produce synthetic polarized dust emission maps for the central $8000 \au$ at a slightly lower resolution than available in the native simulation ($10 \au$ per pixel), at the two wavelengths observations of the dust polarized emission that are the most regularly carried out in current observational facilities ($\lambda = 0.8~\mathrm{mm}$, and  $1.3~\mathrm{mm}$). Note that the dust thermal emission is optically thin at these two wavelengths, throughout most of the spatial scales $r>50~\au$ investigated here (see Fig. \ref{profile_opacity}).

%We use a slightly lower resolution than available in the native simulation ($10 \au$ per pixel), and produce synthetic polarized dust emission maps for the central $8000 \au$, at the two wavelengths observations of the dust polarized emission that are the most regularly carried out in current observational facilities ($\lambda = 0.8~\mathrm{mm}$, and  $1.3~\mathrm{mm}$). Note that the dust thermal emission is optically thin at these two wavelengths, throughout most of the spatial scales $r>50$ au investigated here (see Fig. \ref{profile_opacity})

%%%%%%%%%%%%%%%%%%%%%%%%%%%%%%%%%%%
%%%%%%%%%%%%%%%%%%%%%%%%%%%%%%%%%%%
%%%%%%%%%%%%%%%%%%%%%%%%%%%%%%%%%%%
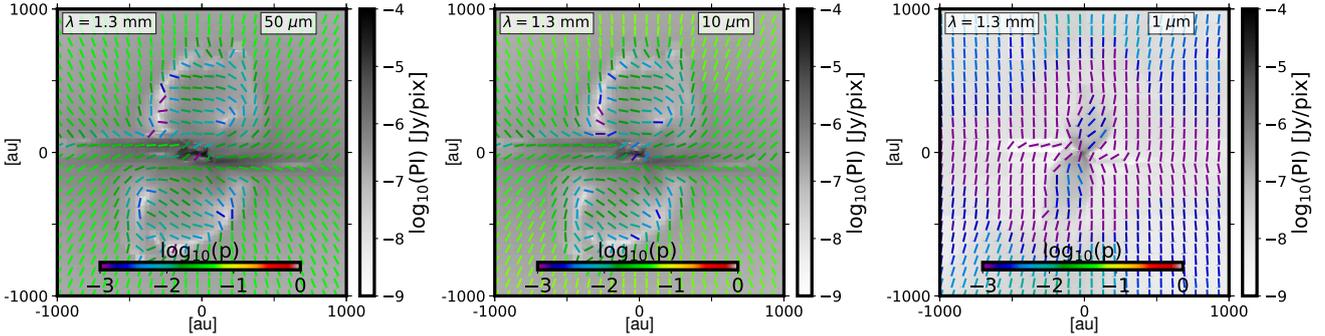
\begin{figure*}
\centering
\hspace{-0.7cm}
\begin{tikzpicture}
\node[above right] (img) at (0,0) {
  \begin{tabular}{@{}lll@{}}
  % trim left, bottom, right, top
  \includegraphics[width=0.325\textwidth, trim={0.75cm 0.65cm 1.3cm 1.2cm},clip]{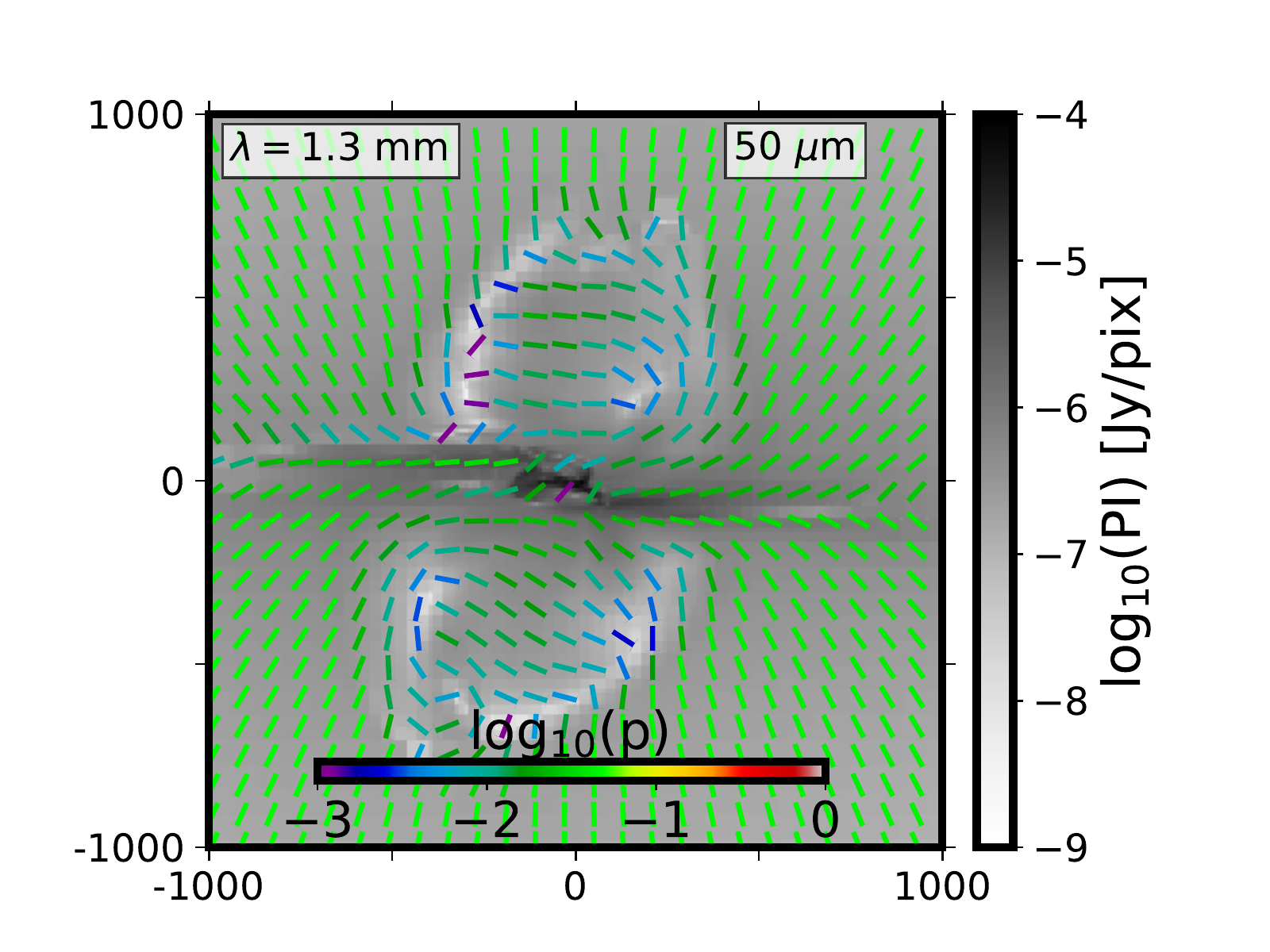}%\\
  \includegraphics[width=0.325\textwidth, trim={0.75cm 0.65cm 1.3cm 1.2cm},clip]{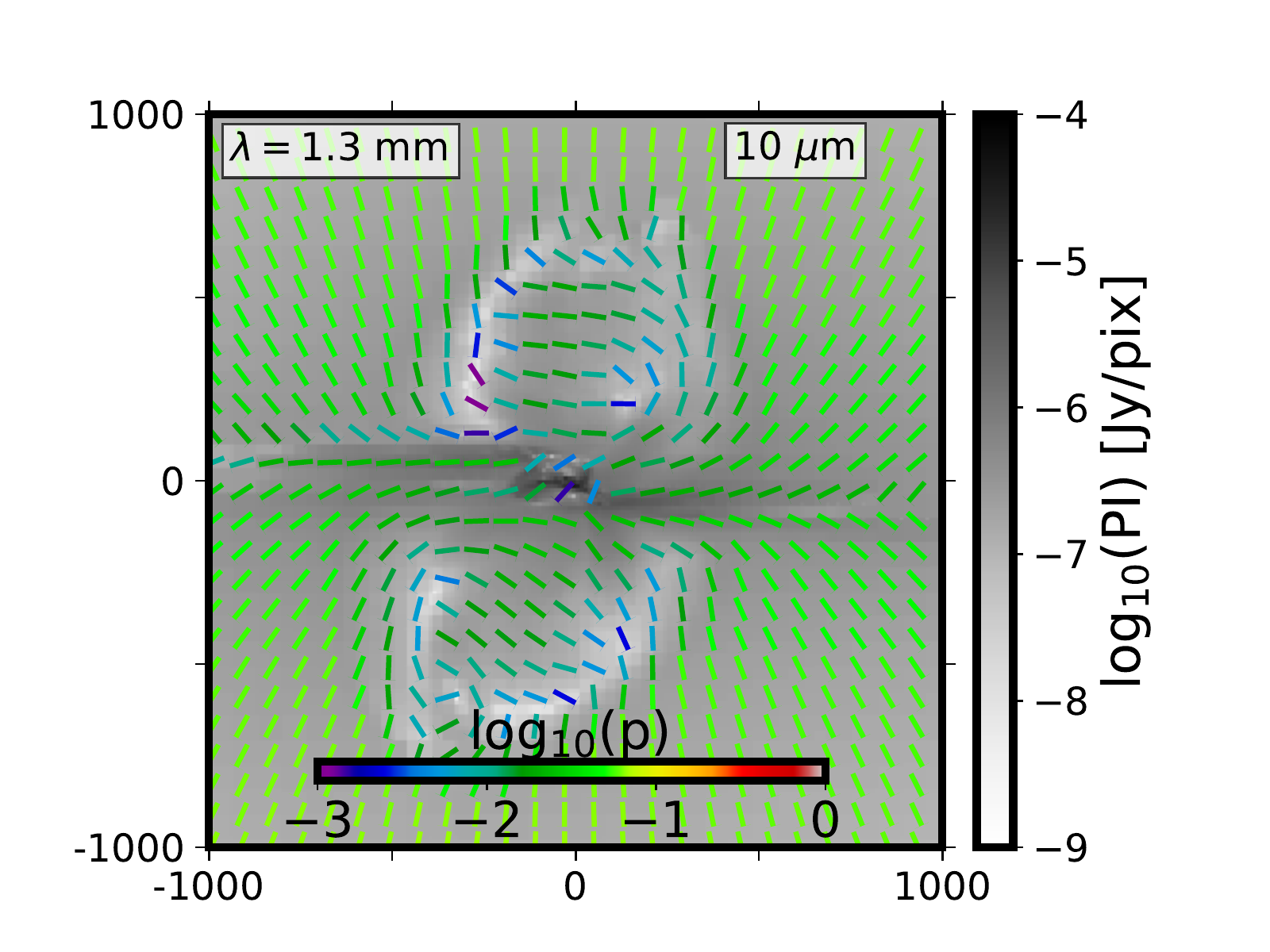} %\\
  \includegraphics[width=0.325\textwidth, trim={0.75cm 0.65cm 1.3cm 1.2cm},clip]{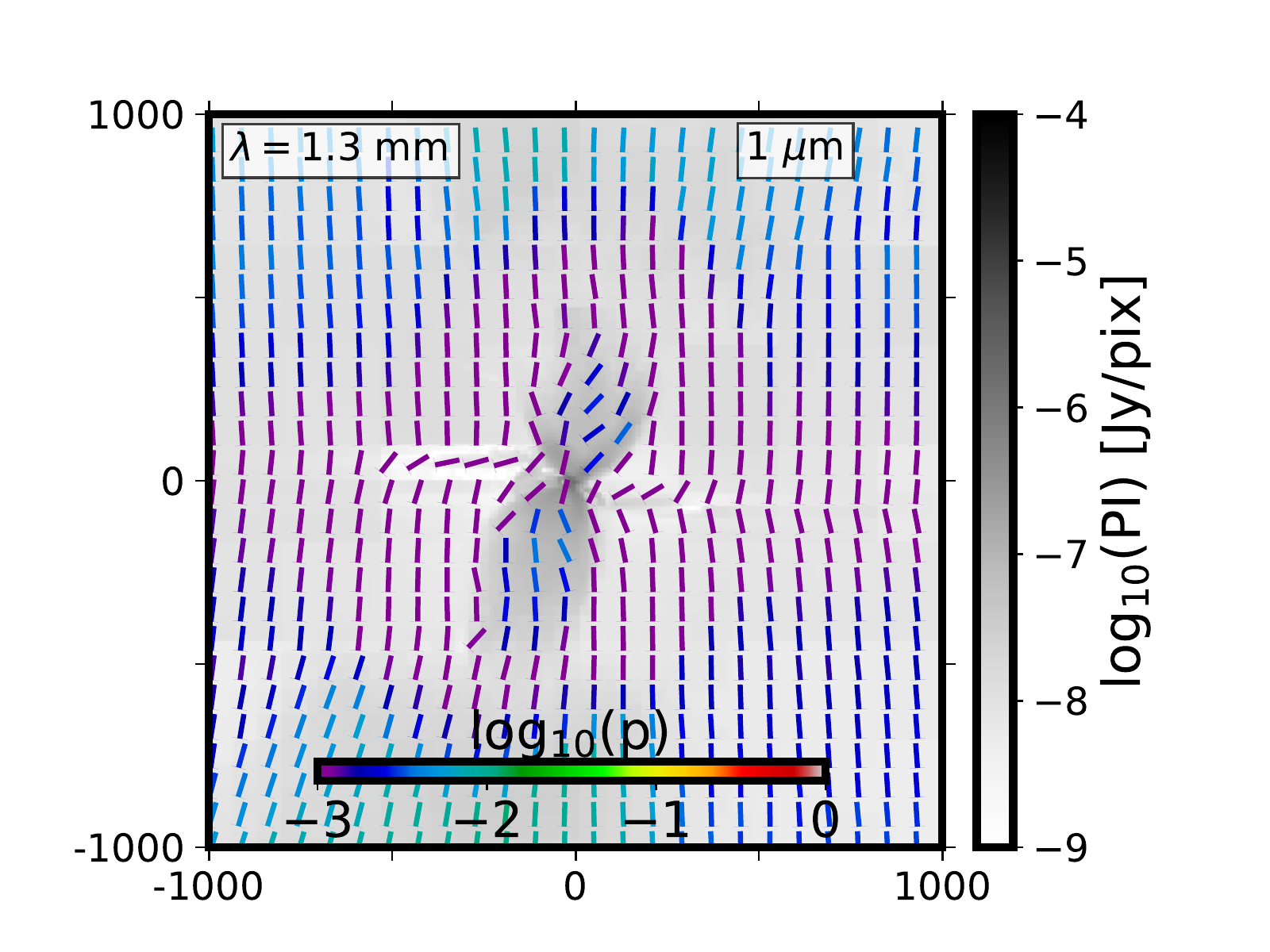}  
  \end{tabular}};
  \node [rotate=90, scale=0.8] at (0.8,2.4) {\fontfamily{phv}\selectfont [au]};
  \node [rotate=90, scale=0.8] at (6.6,2.4) {\fontfamily{phv}\selectfont [au]};
  \node [rotate=90, scale=0.8] at (12.4,2.4) {\fontfamily{phv}\selectfont [au]};
  \node [rotate=0, scale=0.8] at (3.3,0.1) {\fontfamily{phv}\selectfont [au]};
  \node [rotate=0, scale=0.8] at (9.05,0.1) {\fontfamily{phv}\selectfont [au]};
  \node [rotate=0, scale=0.8] at (14.9,0.1) {\fontfamily{phv}\selectfont [au]};
\end{tikzpicture}
\vspace{-0.6cm}
\caption{Perfect synthetic observations at $\lambda=1.3~\mathrm{mm}$ showing the polarized emission (background image), and the inferred magnetic field orientation line segments, for which we have color coded the polarization fraction. $a_\mathrm{max}$ is given on the top-right corner of each panel. The full set of dust distributions is shown in Fig.~\ref{Full_synth}}.
\label{synth}
\end{figure*}

%We model the central source as a blackbody of $1~\mathrm{L_\odot}$. Even though YSOs are deeply embedded in their parent clouds and shielded from the interstellar radiation field (ISRF), in an academic interest we include its presence as an external radiation field to increase the efficiency of the alignment by the RAT mechanism. These two sources are used to heat the dust and participate in the alignment.  
%\begin{comment}
%We have check our results with and without the ISRF and they do not differ too much.
%\end{comment}
%%%%%%%%%%%%%%%%%%%%%%%%%%%%

%We have carried out radiative transfer calculations using 7 different dust grain size distributions as described previously in this section. They are characterized by the maximum dust size in the distribution $a_\mathrm{max}=1, 2, 3, 5, 10, 30,$ and $50~\micron$. The results are discussed in the following section.
%
%We compute the total linearly polarized emission, or polarized intensity ($PI$), and the polarization fraction ($p$) from the Stokes parameters $I$, $Q$ and $U$ as $PI = (Q^2 +U^2)^{0.5}$, and $p = PI/I$.
%The magnetic field orientation is computed by rotating by $90^\circ$ the polarization angle $\psi$ obtained as:  $\psi = \frac{1}{2}\arctan (Q, U)$. 
%
%__________________________________________________________________
%
%\section{Synthetic polarized dust emission maps from solar-type protostellar cores} \label{results}

\textsc{POLARIS} produces maps of the Stokes parameters $I$, $Q$ and $U$, from which we build perfect (not including any instrumental effect) total linearly polarized emission maps (polarized intensity $PI = (Q^2 +U^2)^{0.5}$), and polarization fraction ($p = PI/I$) maps. In Valdivia et al. (in prep), the instrumental effects have been included and show not to affect the polarization fractions by more than $50\%$ in regions where the total dust thermal emission (Stokes I) is detected with a signal-to-noise ratio $>5$ (standard debiasing).
Finally, we produce synthetic magnetic field `vectors' maps by rotating by $90^\circ$ the polarization angle $\psi$ obtained as:  $\psi = \frac{1}{2}\arctan (U, Q)$. 

\begin{figure*}
\centering
  \begin{tabular}{@{}lll@{}}
% trim left, bottom, right, top
    \includegraphics[width=0.32\textwidth, trim={3.5cm 2.4cm 6.05cm 1.2cm},clip]{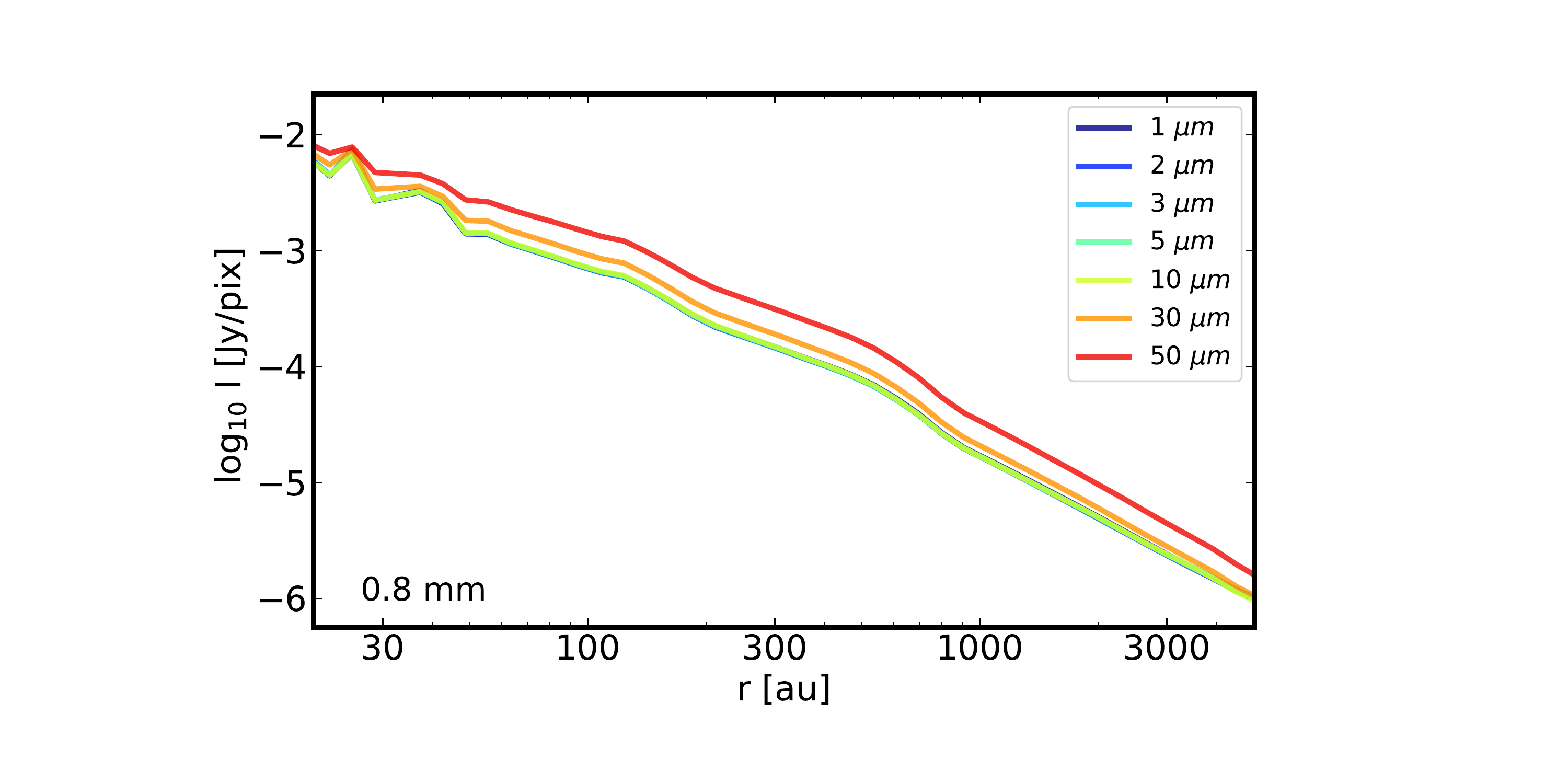}&
    \includegraphics[width=0.32\textwidth, trim={3.5cm 2.4cm 6.05cm 1.2cm},clip]{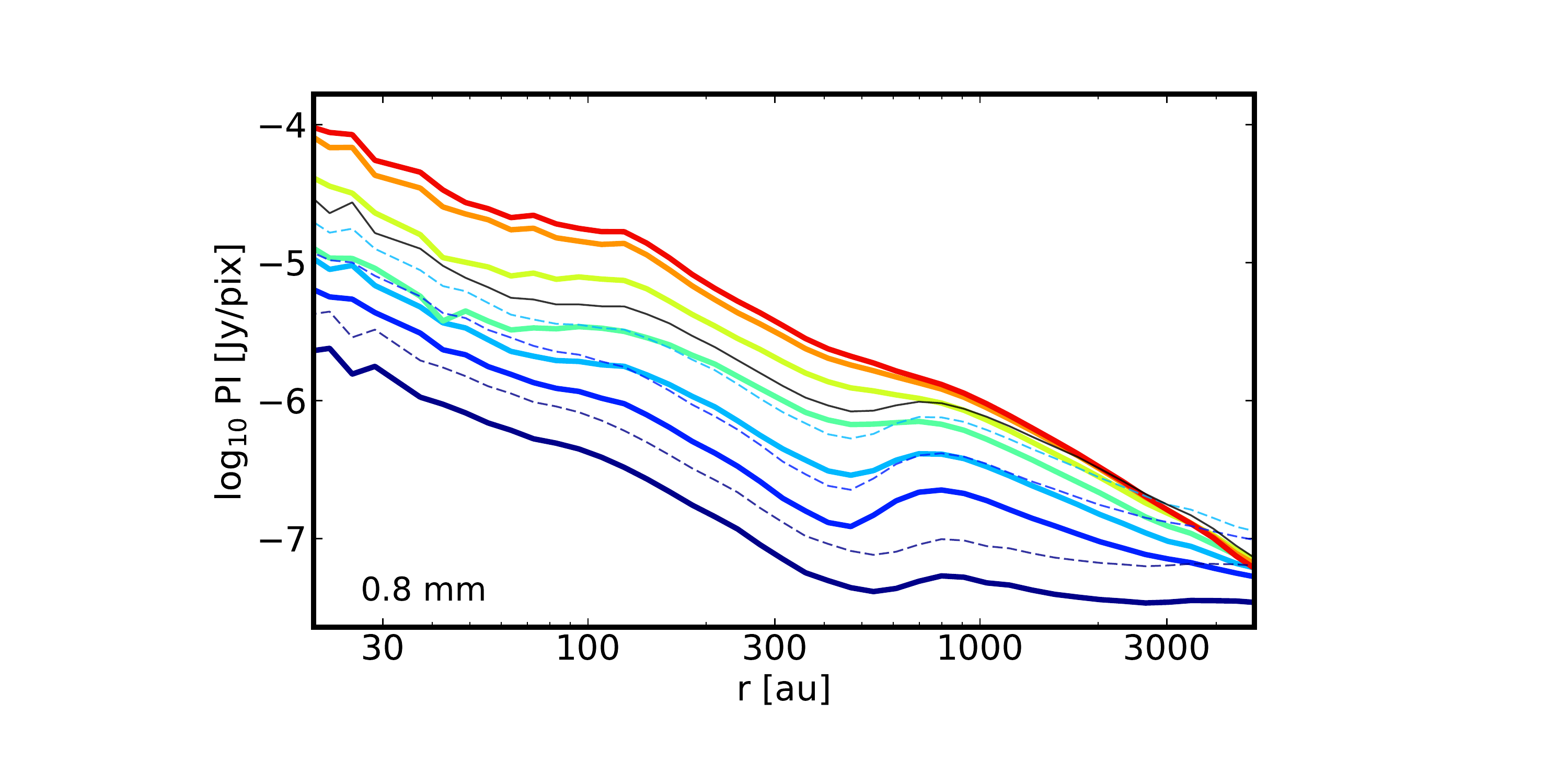}&
    \includegraphics[width=0.32\textwidth, trim={3.5cm 2.4cm 6.05cm 1.2cm},clip]{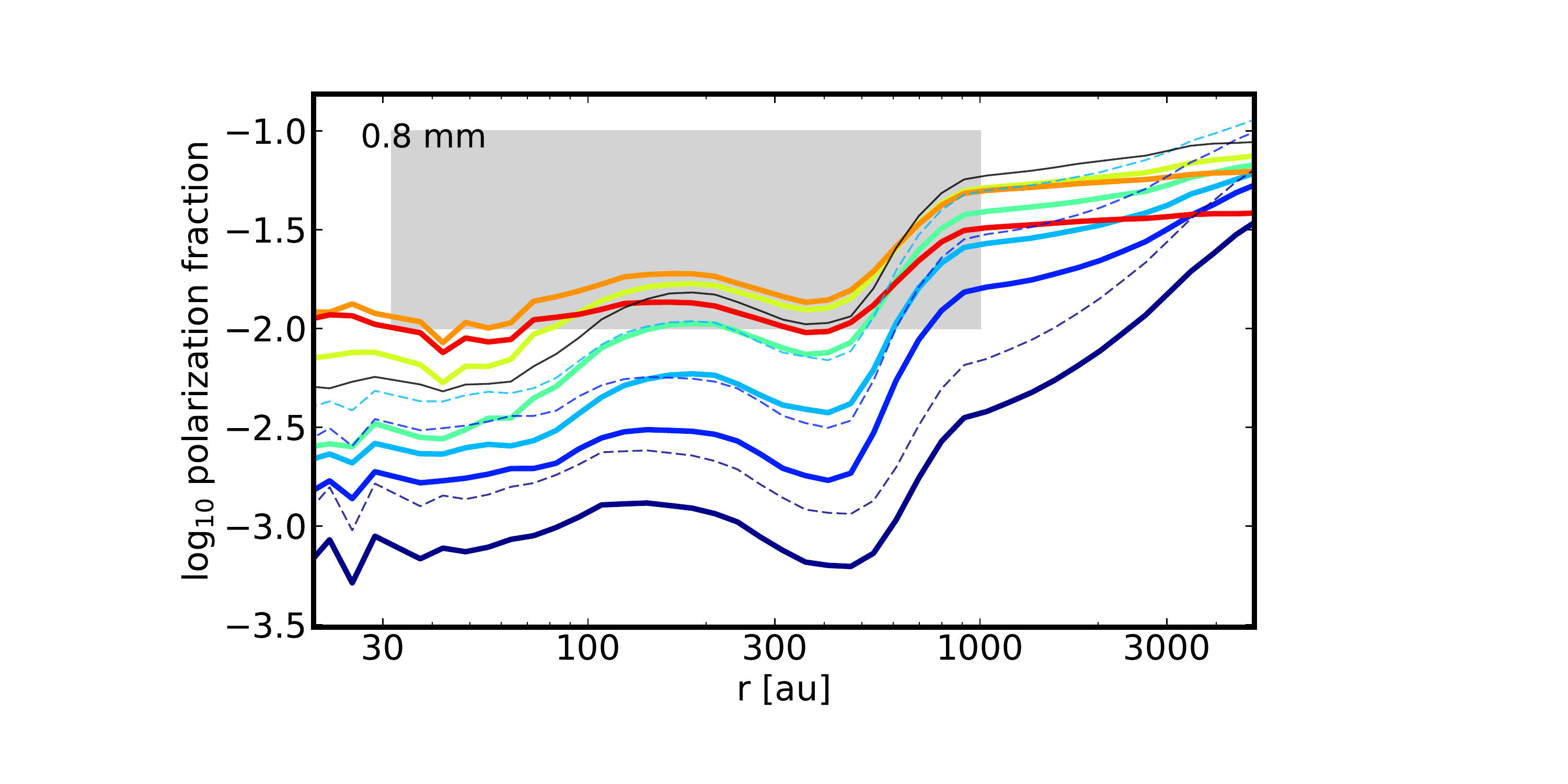}\\
    \includegraphics[width=0.32\textwidth, trim={3.5cm 1.5cm 6.05cm 1.6cm},clip]{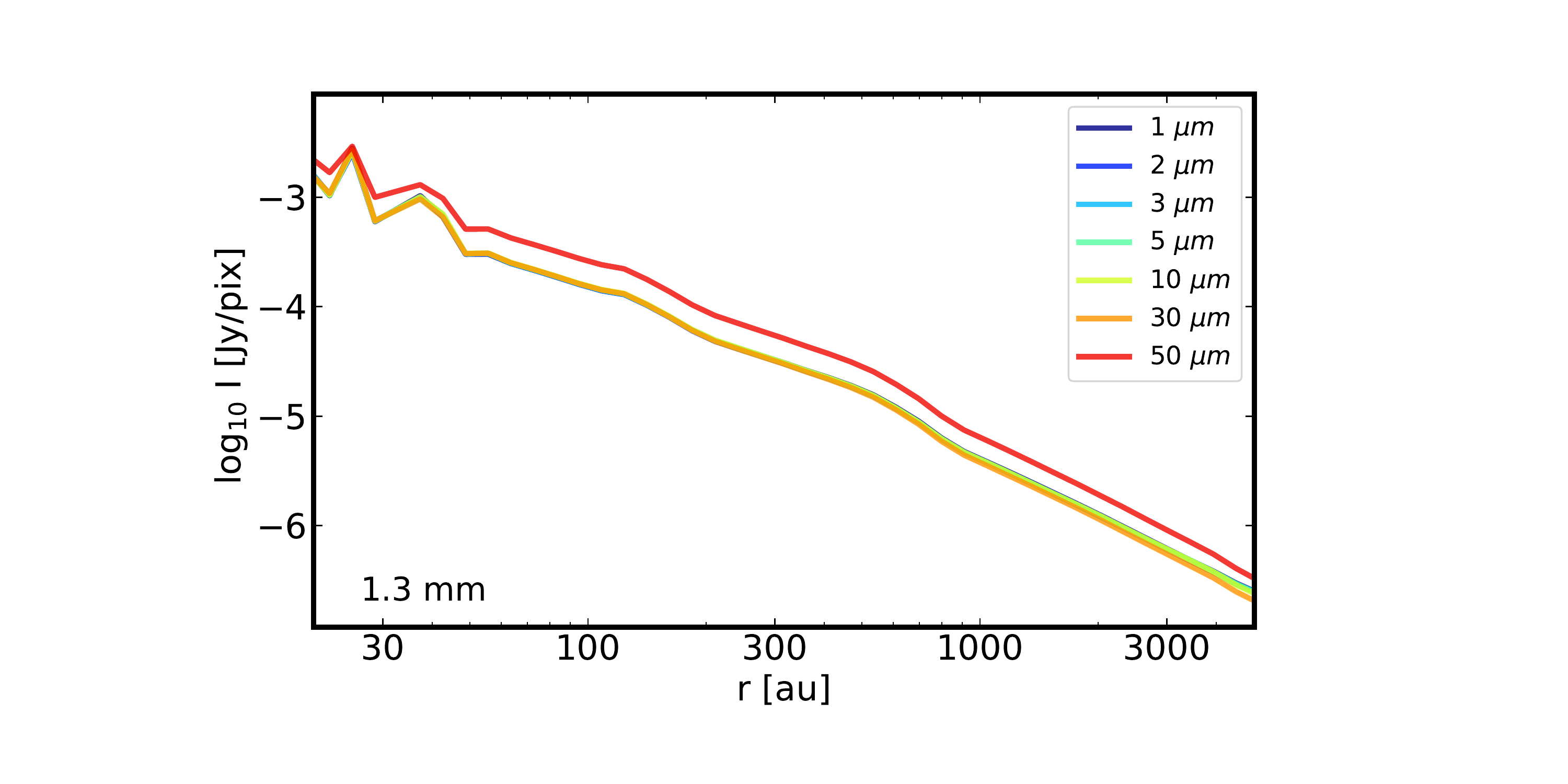}&
    \includegraphics[width=0.32\textwidth, trim={3.5cm 1.5cm 6.05cm 1.2cm},clip]{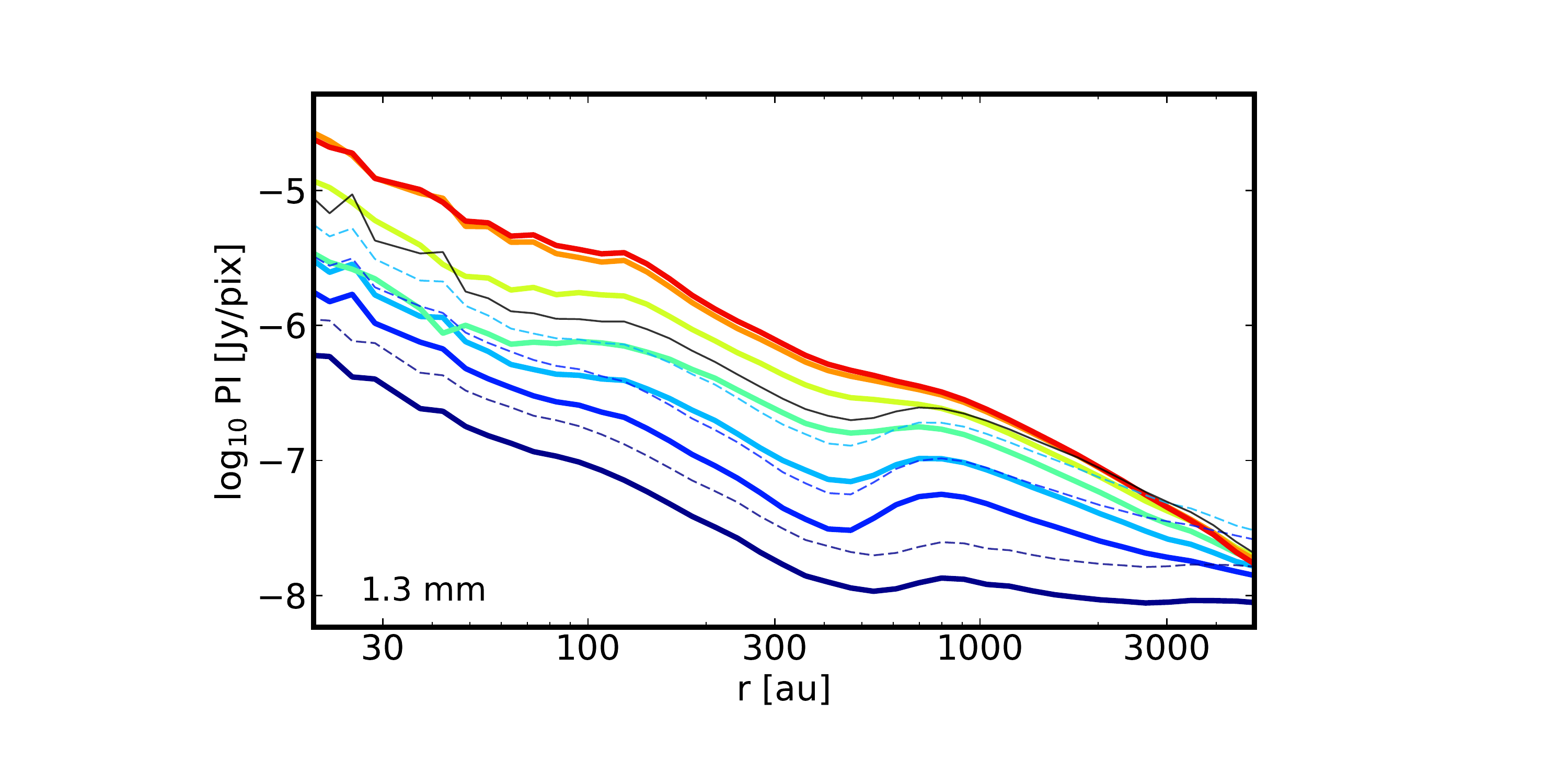}&
    \includegraphics[width=0.32\textwidth, trim={3.5cm 1.5cm 6.05cm 1.2cm},clip]{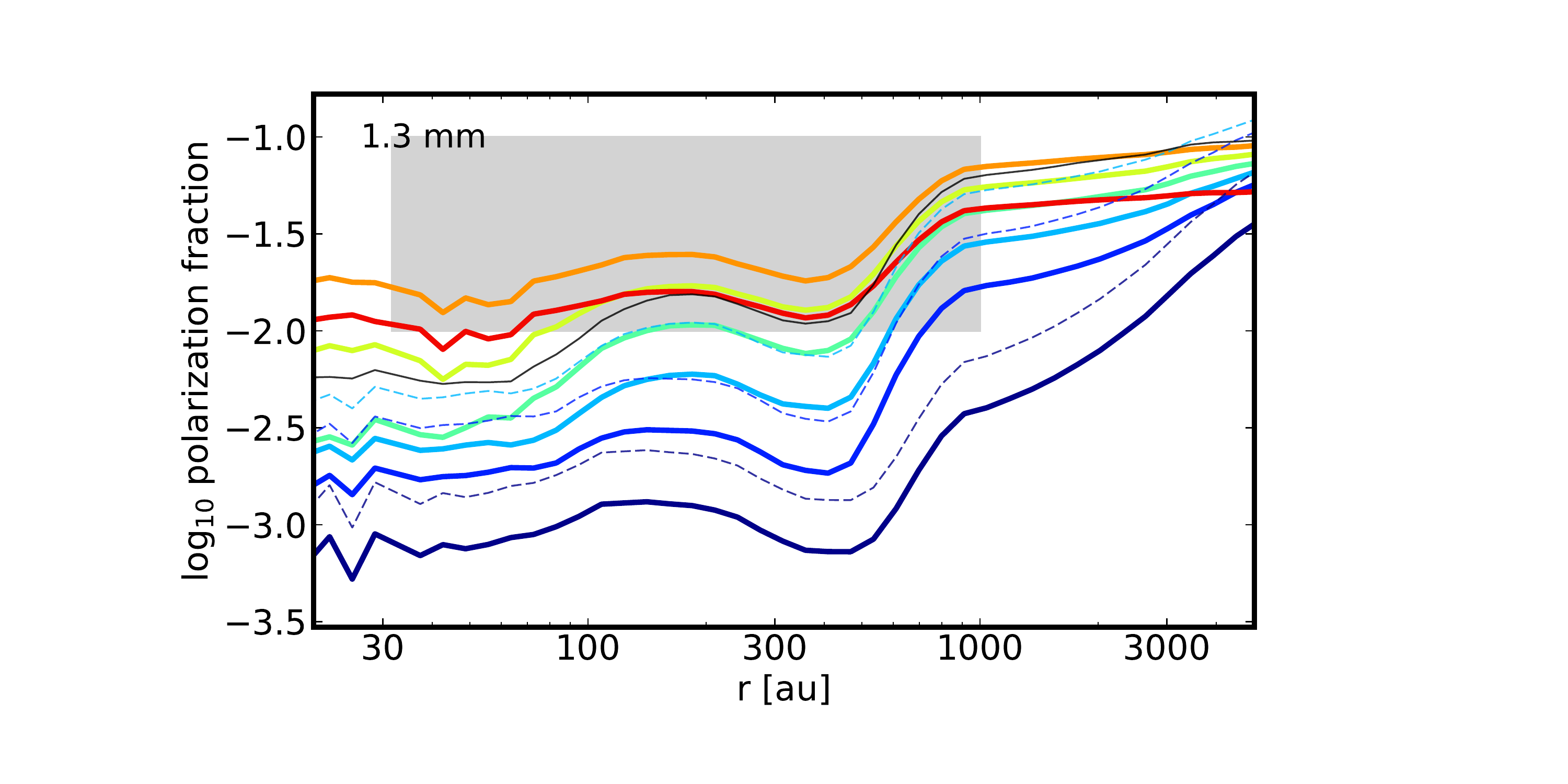}%\\

  \end{tabular}
  \vspace{-0.3cm}
\caption{Radial mean profiles for two wavelengths: $\lambda = 0.8~\mathrm{mm}$ (top panels), and $\lambda = 1.3~\mathrm{mm}$ (bottom panels). From left to right: total dust emission (Stokes I), total linearly polarized intensity ($PI$), and polarization fraction indicating the typical values observed in YSOs (grey region). The dashed lines show the expected values assuming $f_{\mathrm{high}-J}=1$. The thin solid black line shows the polarization with $a_\mathrm{max}=3~\micron$ and an unlikely size distribution of $n(a) \propto a^{-2}~\mathrm{d}a $, for reference.}
\label{radial}
\end{figure*}

Figure~\ref{synth} shows the resulting polarized dust continuum emission maps at $\lambda = 1.3~\mathrm{mm}$, overlaid with the magnetic field orientation line segments reconstructed from the Stokes maps, for which we have color coded the recovered polarization fraction. We show the resulting maps for three dust grain size distributions, zoomed in the central $2000~\mathrm{au}$ of the original map so the features in the inner envelopes can be easily distinguished.
This figure shows that the resulting polarized emission strongly depends on the dust grain sizes (see the $a_\mathrm{alig}$ map in Fig.~\ref{aalig}), and especially on the maximum dust grain size of the dust distribution included in the radiative transfer. This can be easily understood since the radiative torque applied to dust grains $Q$ depends strongly on the ratio between the incoming photon wavelength and the dust grain size \citep{LazarianHoang07, Hoang14}. 
%In particular, this figure shows that grains smaller than $1$ or $2 ~\micron$ can only be aligned in the irradiated cavities created by the bipolar outflows (bipolar feature seen in emission in the lower panel), where photons from the accreting protostar can travel more freely: the `$\textit{\textbf{B}}-$vectors' inferred from these grains trace the very local magnetic field direction in the outflow, but if no larger grains are included in the model the resulting polarization fraction is very low ($p <1\%$), everywhere in the map.
In particular, this figure shows that grains smaller than $1$ or $2 ~\micron$ can only be aligned in the irradiated cavities created by the bipolar outflows (bipolar feature seen in emission in the lower panel), where photons from the accreting protostar can travel more freely. With such grain size distributions, the resulting polarization fraction on the lines of sight probing the envelope remain very low ($p <1\%$), even for a pure silicate composition (see Fig.~\ref{Silicates}).
Only when including dust grains with sizes up to $10 ~\micron$, the envelope polarized emission is recovered from the high density equatorial region at polarization fractions $p>1\%$.\\
\indent In Fig.~\ref{radial} we present the radial profiles of the total dust thermal emission (left panels), polarized intensity (middle panels) and polarization fraction (right panels), for the  dust grain size distributions explored, at the two wavelengths $\lambda = 0.8~\mathrm{mm}$ (top row) and  $\lambda = 1.3~\mathrm{mm}$ (bottom row). 
The dust thermal emission shows a decrease towards the outer regions of the envelope, consistent with the decrease of the column densities and dust temperature, which remains quite insensitive to the dust size distribution (see Fig.~\ref{profile_temp}). 
Even though the dust size distribution has a mild influence on the total dust emission, the center and right panels show that the effect on its polarized component is very strong, producing a variation of the polarization fraction of roughly two orders of magnitude in the inner envelope ($r<1000~\au$, and densities higher than $10^7~\mathrm{cm^{-3}}$). 
%The polarized intensity increases with $a_\mathrm{max}$ until $a_\mathrm{max}\sim 30~\micron$. This is more striking in the case of the polarization fraction, where the distribution with $a_\mathrm{max}\sim 50~\micron$ shows a lower polarization fraction. Since the total dust mass is kept constant, the cross section per unit mass decreases with $a_\mathrm{max}$, and as the efficiency of absorption and emission do not increase further for distributions with larger grains, the contribution to the polarized emission decreases.
The polarized intensity increases with $a_\mathrm{max}$ until $a_\mathrm{max}\sim 30~\micron$. Since the total dust mass is kept constant, the cross section per unit mass decreases with $a_\mathrm{max}$, and as the efficiency of absorption and emission do not increase further for distributions with larger grains, the contribution to the polarized emission starts to decrease. This is more striking in the case of the polarization fraction, where the distribution with $a_\mathrm{max}\sim 50~\micron$ shows a lower polarization fraction. 

%distributions with larger grains do not further contribute to the polarized emission.  
%This is more striking in the case of the polarization fraction, where there the distribution with $a_\mathrm{max}\sim 50~\micron$ shows a lower polarization fraction. Several \textcolor{red}{things} can be contributing: since the total dust mass is kept constant, the number density of dust grains able to align decreases for $a_\mathrm{max}>10~\micron$. 
%Other possibility is that grains larger than $1~\micron$ can align either parallel of perpendicular to the magnetic field lines \citep{Lazarian2011}, leading to a lower polarization degree. 
%We disregard the intrinsic inability of big grains to produce a polarized emission since all the dust size distributions are below the geometrical regime \citep{Cho2007} for the observed wavelengths \textcolor{green}{AM: this sentence needs to be clarified, or removed (too specific)}.     

\section{Polarized dust emission as an indicator of early grain growth in young protostars?}\label{discussion}

%Even though polarization observations, both in emission and extinction, point towards variations in the grain properties when the gas density in the ISM increases \citep{Juvela2018}, 
It is only in the last decade that observations of the polarized dust emission in the inner regions of low-mass protostellar cores, at scales $\lesssim 1000\au$ could be carried out, starting with the advent of polarization capabilities on the Submillimeter Array \citep[SMA, ][]{Girart2006, Goncalves2008, Frau2011, Rao2009, Galametz2018} and the Combined Array for Research in Millimeter-wave Astronomy \citep[CARMA, ][]{Stephens2013, Hull2014, Segura-Cox2015}. 
Since 2016 the most sensitive sub-millimeter interferometer, ALMA, is producing exquisite polarized dust continuum emission maps from the high-density protostellar interiors, probing the properties of polarized emission down to scales of a few tens of $\au$. Typically, polarization fractions of $p \sim 2-10\%$ are recovered in envelopes at radii $>50-5000~\au$, when both self-scattering polarization and large filtering effects can be excluded \citep[e.g., ][]{Kataoka2015, Hull2017, Maury2018, Cox2018, Alves2018, Sadavoy2018}.

All dust grain size distributions including only small grains (e.g. $<3~\micron$) considered in our study produce polarized emission intensities at undetectable levels with current interferometers, and polarization fractions too low by one order of magnitude with respect to observed values at similar scales in solar-type protostars (see grey region in Fig.~\ref{radial}). Indeed, the local radiation field is drastically shifted towards longer wavelengths in the high density interiors of protostellar envelopes, and since only dust grains with sizes close to the wavelength of incident photons get efficiently aligned, our models require large grains to produce polarized emission from protostellar cores at detectable levels. Furthermore, in regions with high rates of collisions, such as high pressure regions in the equatorial plane, larger grains remain aligned more easily due to their higher inertia.
We stress that a harder radiation field from a higher accretion luminosity is not able to significantly increase the polarization fraction for these grain sizes (see Fig.~\ref{radial10Lsun}). 
If RATs are indeed responsible for the grain alignment in dense protostellar interiors, then our results suggest that protostellar envelopes must include dust grains with typical sizes of at least $10 ~\micron$ to produce polarization fractions similar to the observed values $p>>1\%$. 

Note that, inversely, a lack of detected polarization does not suggest a lack of large grains, but can be due to various factors, such as a beam dilution or line-of-sight canceling due to turbulent or complex magnetic field topologies. Indeed, single dish observations of the dust polarized emission, probing $\sim 0.1 \pc$ scales, show a decrease in polarization fraction in denser regions $p \propto I^{-\alpha}$, with $\alpha \sim 0.5-1.2$ is usually found at those scales \citep{Wolf2003,Crutcher2004,Soam2018}. High angular resolution observations suggest this is at least partly due to a beam averaging effect (unresolved structure of $\textit{\textbf{B}}$) since larger polarization fractions are recovered when probing typical scales $1000 \au$ in dense cores, thanks to interferometric facilities \citep{Hull2014, Hull2017b, Galametz2018}. This was also suggested by early analysis of synthetic polarization maps from MHD simulations \citep{Falceta-Goncalves2008, Padovani2012}. 
In the case of B335, for example, the polarization fraction decreases with the Stokes I intensity in the seven JCMT POL-2 detections \cite[about $\sim 1\%$ on average in the inner 3000 au, see][]{Yen2019}, but is recovered at the level of $\sim 10\%$ with SMA at scales $\sim 500-1000~\au$ \citep{Galametz2018} and with ALMA at similar levels down to scales $\sim 50-100$ au \citep{Maury2018}.

Although observational studies suggest that the RATs is likely to be a dominant mechanism of grain alignment in the ISM, we acknowledge that developments regarding alignment theories are thriving and that currently the knowledge is still limited \citep[see][for a discussion]{Lazarian2009}. Especially, we note that a new flavor of RATs including super-paramagnetic grains \citep[MRAT, ][]{Hoang2016}, able to increase the fraction of aligned grains in the high-$J$ attractor, was developed to reproduce the high levels of polarized emission observed in the diffuse ISM by the {\it{Planck}} observatory, $p$ up to $\sim 20\%$ \citep{PlanckXXI2015}. Note that in case a mechanism would allow to align all dust grains, independently of their size with respect to the local radiation field (perfect alignment, see Fig. \ref{radialPA}), 
then no constraints could be obtained on the dust size distribution based on the polarization fractions. However, such a mechanism is yet unconstrained: our results are valid in the current widely admitted framework of RATs dominated alignment, and should be revised if other mechanisms are shown to produce as good agreement as RATs with observations in the future. 
A major caveat in the RATs theory is the actual value of the fraction of dust spinning at suprathermal velocities in the high-$J$ attractor ($f_{\mathrm{high-}J}$), that corresponds to a highly stable population and can be considered  as perfectly aligned. In our study we use a fraction $f_{\mathrm{high-}J} = 0.25$, that fits better the polarization fractions observed in Bok globules \citep{Brauer16}. 
Additionally, we give an upper limit to the polarized intensity and to the polarization fraction for the distributions with $a_\mathrm{max} \leq 3 ~\micron$ by assuming a rather unrealistic fraction of $f_{\mathrm{high-}J} = 1.0$, shown in dashed lines in the same figure.
Note that including the ISRF slightly enhances the alignment by the RATs mechanism only in the outer envelope layers where the ISRF photons are able to penetrate (up  to $5\times 10^{-2}$ in polarization fraction for $r>1000 \au$). It has a very low impact on the polarization fraction in the inner $1000 \au$ (typically less than $5\times 10^{-3}$ in polarization fraction).
%Finally, another issue is the slope of the dust size distribution. We include for reference the results for $PI$ and $p$ for the $a_\mathrm{max} = 3 ~\micron$ distribution using a very unlikely slope of $-2$ (thin solid black line). This two assumptions show that dust size has a stronger influence on the polarization fraction,and that, given the uncertainties in the slope of the dust size distribution and on the fraction of dust grains in the high-$J$ state, size distributions with only small grains ($a<3 ~\micron$) can barely reach the polarization levels of distributions that include dust grains of sizes $a\sim 5 ~\micron$.
Another possible caveat is the slope of the grain size distribution. We include for reference the results for $PI$ and $p$ for the $a_\mathrm{max} = 3 ~\micron$ distribution using an unlikely slope of $-2$ (thin solid black line). Altogether, Fig.~\ref{radial} shows that realistic dust size distributions with only small grains ($a<3 ~\micron$) can barely reach the observed degrees of polarization. 
Additionally, we acknowledge the difficulty to perform precise calculations for large dust grains consisting of complex aggregates \citep{Min2009}, and that the optical constants and cross sections used in grain models might not be fully accurate and need to be reassessed in the near future, at the light of the newest laboratory measurements on analogues of interstellar silicates \citep{Demyk2017a, Demyk2017b}.
Finally, although grain growth may make grains more spherical and thus harder to align with $B$-fields by radiative torques, the polarization fractions observed in, e.g., the inner envelopes of Class 0 protostars cannot be explained with diffuse ISM grains only in the currently developed alignment theories. 
%\textcolor{green}{AM: Here Add one or two sentences to say that we tried to increase the central source luminosity, and let the simulation evolve twice longer and it does not change significantly the results. Mention we recover the right Stokes I flux compared to B335 (refer to ALMA synthetic observations in Valdivia et al. in prep) ?}
%We also stress that a lack of detected polarization does not always points toward a lack of grains with sizes adapted to the local radiation field so they can be aligned

%\textcolor{red}{Is perfect alignment compatible with decrease of polar fraction seen from IR to mm wavelengths ? Link with B structure (larger dispersion at higher densities) cannot be discarded?}

\section{Conclusions}

We explore radiative transfer maps of the polarized dust continuum emission from a synthetic low-mass protostellar core. We show that, in the RATs paradigm of dust grain alignment with the magnetic field, only radiative transfer models including large grains (typically $>10 ~\micron$) are able to produce polarized dust emission at levels similar to what is currently observed at small scales within embedded protostellar cores in the (sub-)mm domain (e.g. ALMA).
Our results suggest that either grains have already significantly grown at scales $100-1000 \au$ in the youngest protostellar envelopes observed $<10^5$ years after the onset of collapse, or a missing alignment theory remains to be developed that would virtually align perfectly all dust grains, and this independently of their size with respect to the wavelength of the local radiation field.
Considering the current theoretical difficulties to align ISM type nanometric grains at the high gas densities typical in these protostars, our work favors the possibility of grain growth and highlights the possibility to use the properties of dust polarized emission to gain access to dust pristine properties and better describe, for example, the very early stages of the formation of planetesimals.

%__________________________________________________________________
%__________________________________________________________________
%__________________________________________________________________
%__________________________________________________________________
%__________________________________________________________________
%__________________________________________________________________
%__________________________________________________________________
%__________________________________________________________________

\section*{Acknowledgements}

This research has received funding from the European Research Council (ERC) under the European Union's Horizon 2020 research and innovation programme (MagneticYSOS project, Grant Agreement No 679937).\\
This work was supported by the Programme National ``Physique et Chimie du Milieu Interstellaire'' (PCMI) of CNRS/INSU with INC/INP co-funded by CEA and CNES.

%%%%%%%%%%%%%%%%%%%%%%%%%%%%%%%%%%%%%%%%%%%%%%%%%%



%\bibliographystyle{mnras}

\bibliographystyle{mnras} % style aa.bst 
\bibliography{biblio_letter} % your references Yourfile.bib

\begin{thebibliography}{}
\makeatletter
\relax
\def\mn@urlcharsother{\let\do\@makeother \do\$\do\&\do\#\do\^\do\_\do\%\do\~}
\def\mn@doi{\begingroup\mn@urlcharsother \@ifnextchar [ {\mn@doi@}
  {\mn@doi@[]}}
\def\mn@doi@[#1]#2{\def\@tempa{#1}\ifx\@tempa\@empty \href
  {http://dx.doi.org/#2} {doi:#2}\else \href {http://dx.doi.org/#2} {#1}\fi
  \endgroup}
\def\mn@eprint#1#2{\mn@eprint@#1:#2::\@nil}
\def\mn@eprint@arXiv#1{\href {http://arxiv.org/abs/#1} {{\tt arXiv:#1}}}
\def\mn@eprint@dblp#1{\href {http://dblp.uni-trier.de/rec/bibtex/#1.xml}
  {dblp:#1}}
\def\mn@eprint@#1:#2:#3:#4\@nil{\def\@tempa {#1}\def\@tempb {#2}\def\@tempc
  {#3}\ifx \@tempc \@empty \let \@tempc \@tempb \let \@tempb \@tempa \fi \ifx
  \@tempb \@empty \def\@tempb {arXiv}\fi \@ifundefined
  {mn@eprint@\@tempb}{\@tempb:\@tempc}{\expandafter \expandafter \csname
  mn@eprint@\@tempb\endcsname \expandafter{\@tempc}}}

\bibitem[\protect\citeauthoryear{{Alves} et~al.,}{{Alves}
  et~al.}{2018}]{Alves2018}
{Alves} F.~O.,  et~al., 2018, \mn@doi [\aap] {10.1051/0004-6361/201832935},
  \href {http://adsabs.harvard.edu/abs/2018A%26A...616A..56A} {616, A56}

\bibitem[\protect\citeauthoryear{{Andersson}, {Lazarian}  \&
  {Vaillancourt}}{{Andersson} et~al.}{2015}]{Andersson2015}
{Andersson} B.-G.,  {Lazarian} A.,   {Vaillancourt} J.~E.,  2015, \mn@doi
  [\araa] {10.1146/annurev-astro-082214-122414}, \href
  {http://adsabs.harvard.edu/abs/2015ARA%26A..53..501A} {53, 501}

\bibitem[\protect\citeauthoryear{{Andr\'e}, {Ward-Thompson}  \&
  {Barsony}}{{Andr\'e} et~al.}{2000}]{Andre2000}
{Andr\'e} P.,  {Ward-Thompson} D.,   {Barsony} M.,  2000, Protostars and
  Planets IV, \href {http://adsabs.harvard.edu/abs/2000prpl.conf...59A} {p.~59}

\bibitem[\protect\citeauthoryear{{Birnstiel}, {Fang}  \&
  {Johansen}}{{Birnstiel} et~al.}{2016}]{Birnstiel2016}
{Birnstiel} T.,  {Fang} M.,   {Johansen} A.,  2016, \mn@doi [\ssr]
  {10.1007/s11214-016-0256-1}, \href
  {https://ui.adsabs.harvard.edu/\#abs/2016SSRv..205...41B} {205, 41}

\bibitem[\protect\citeauthoryear{{Blum} \& {Wurm}}{{Blum} \&
  {Wurm}}{2008}]{Blum2008}
{Blum} J.,  {Wurm} G.,  2008, \mn@doi [\araa]
  {10.1146/annurev.astro.46.060407.145152}, \href
  {http://adsabs.harvard.edu/abs/2008ARA%26A..46...21B} {46, 21}

\bibitem[\protect\citeauthoryear{{Bracco} et~al.,}{{Bracco}
  et~al.}{2017}]{Bracco2017}
{Bracco} A.,  et~al., 2017, \mn@doi [\aap] {10.1051/0004-6361/201731117}, \href
  {https://ui.adsabs.harvard.edu/\#abs/2017A&A...604A..52B} {604, A52}

\bibitem[\protect\citeauthoryear{{Brauer}, {Wolf}  \& {Reissl}}{{Brauer}
  et~al.}{2016}]{Brauer16}
{Brauer} R.,  {Wolf} S.,   {Reissl} S.,  2016, \mn@doi [\aap]
  {10.1051/0004-6361/201527546}, \href
  {http://adsabs.harvard.edu/abs/2016A%26A...588A.129B} {588, A129}

\bibitem[\protect\citeauthoryear{{Chac{\'o}n-Tanarro}, {Caselli}, {Bizzocchi},
  {Pineda}, {Harju}, {Spaans}  \& {D{\'e}sert}}{{Chac{\'o}n-Tanarro}
  et~al.}{2017}]{ChaconTanarro2017}
{Chac{\'o}n-Tanarro} A.,  {Caselli} P.,  {Bizzocchi} L.,  {Pineda} J.~E.,
  {Harju} J.,  {Spaans} M.,   {D{\'e}sert} F.~X.,  2017, \mn@doi [\aap]
  {10.1051/0004-6361/201630265}, \href
  {https://ui.adsabs.harvard.edu/\#abs/2017A&A...606A.142C} {606, A142}

\bibitem[\protect\citeauthoryear{{Cho} \& {Lazarian}}{{Cho} \&
  {Lazarian}}{2007}]{Cho2007}
{Cho} J.,  {Lazarian} A.,  2007, \mn@doi [\apj] {10.1086/521805}, \href
  {http://adsabs.harvard.edu/abs/2007ApJ...669.1085C} {669, 1085}

\bibitem[\protect\citeauthoryear{{Cox}, {Harris}, {Looney}, {Li}, {Yang},
  {Tobin}  \& {Stephens}}{{Cox} et~al.}{2018}]{Cox2018}
{Cox} E.~G.,  {Harris} R.~J.,  {Looney} L.~W.,  {Li} Z.-Y.,  {Yang} H.,
  {Tobin} J.~J.,   {Stephens} I.,  2018, \mn@doi [\apj]
  {10.3847/1538-4357/aaacd2}, \href
  {http://adsabs.harvard.edu/abs/2018ApJ...855...92C} {855, 92}

\bibitem[\protect\citeauthoryear{{Crutcher}}{{Crutcher}}{2004}]{Crutcher2004}
{Crutcher} R.~M.,  2004, \mn@doi [\apss] {10.1023/B:ASTR.0000045021.42255.95},
  \href {http://adsabs.harvard.edu/abs/2004Ap%26SS.292..225C} {292, 225}

\bibitem[\protect\citeauthoryear{{Davis} \& {Greenstein}}{{Davis} \&
  {Greenstein}}{1951}]{Davis51}
{Davis} Jr. L.,  {Greenstein} J.~L.,  1951, \mn@doi [\apj] {10.1086/145464},
  \href {http://adsabs.harvard.edu/abs/1951ApJ...114..206D} {114, 206}

\bibitem[\protect\citeauthoryear{{Demyk} et~al.,}{{Demyk}
  et~al.}{2017a}]{Demyk2017a}
{Demyk} K.,  et~al., 2017a, \mn@doi [\aap] {10.1051/0004-6361/201629711}, \href
  {http://adsabs.harvard.edu/abs/2017A%26A...600A.123D} {600, A123}

\bibitem[\protect\citeauthoryear{{Demyk} et~al.,}{{Demyk}
  et~al.}{2017b}]{Demyk2017b}
{Demyk} K.,  et~al., 2017b, \mn@doi [\aap] {10.1051/0004-6361/201730944}, \href
  {http://adsabs.harvard.edu/abs/2017A%26A...606A..50D} {606, A50}

\bibitem[\protect\citeauthoryear{{Dominik}, {Paszun}  \& {Borel}}{{Dominik}
  et~al.}{2016}]{Dominik2016}
{Dominik} C.,  {Paszun} D.,   {Borel} H.,  2016, arXiv e-prints, \href
  {http://adsabs.harvard.edu/abs/2016arXiv161100167D} {}

\bibitem[\protect\citeauthoryear{{Draine} \& {Weingartner}}{{Draine} \&
  {Weingartner}}{1996}]{Draine1996}
{Draine} B.~T.,  {Weingartner} J.~C.,  1996, \mn@doi [\apj] {10.1086/177887},
  \href {http://adsabs.harvard.edu/abs/1996ApJ...470..551D} {470, 551}

\bibitem[\protect\citeauthoryear{{Draine} \& {Weingartner}}{{Draine} \&
  {Weingartner}}{1997}]{Draine97}
{Draine} B.~T.,  {Weingartner} J.~C.,  1997, \mn@doi [\apj] {10.1086/304008},
  \href {http://adsabs.harvard.edu/abs/1997ApJ...480..633D} {480, 633}

\bibitem[\protect\citeauthoryear{{Dunham} et~al.,}{{Dunham}
  et~al.}{2014}]{Dunham2014}
{Dunham} M.~M.,  et~al., 2014, \mn@doi [Protostars and Planets VI]
  {10.2458/azu_uapress_9780816531240-ch009}, \href
  {http://adsabs.harvard.edu/abs/2014prpl.conf..195D} {p.~195}

\bibitem[\protect\citeauthoryear{{Falceta-Gon{\c c}alves}, {Lazarian}  \&
  {Kowal}}{{Falceta-Gon{\c c}alves} et~al.}{2008}]{Falceta-Goncalves2008}
{Falceta-Gon{\c c}alves} D.,  {Lazarian} A.,   {Kowal} G.,  2008, \mn@doi
  [\apj] {10.1086/587479}, \href
  {http://adsabs.harvard.edu/abs/2008ApJ...679..537F} {679, 537}

\bibitem[\protect\citeauthoryear{{Flagey} et~al.,}{{Flagey}
  et~al.}{2009}]{Flagey2009}
{Flagey} N.,  et~al., 2009, \mn@doi [\apj] {10.1088/0004-637X/701/2/1450},
  \href {http://adsabs.harvard.edu/abs/2009ApJ...701.1450F} {701, 1450}

\bibitem[\protect\citeauthoryear{{Frau}, {Galli}  \& {Girart}}{{Frau}
  et~al.}{2011}]{Frau2011}
{Frau} P.,  {Galli} D.,   {Girart} J.~M.,  2011, \mn@doi [\aap]
  {10.1051/0004-6361/201117813}, \href
  {http://adsabs.harvard.edu/abs/2011A%26A...535A..44F} {535, A44}

\bibitem[\protect\citeauthoryear{{Fromang}, {Hennebelle}  \&
  {Teyssier}}{{Fromang} et~al.}{2006}]{Fromang2006}
{Fromang} S.,  {Hennebelle} P.,   {Teyssier} R.,  2006, \mn@doi [\aap]
  {10.1051/0004-6361:20065371}, \href
  {http://adsabs.harvard.edu/abs/2006A%26A...457..371F} {457, 371}

\bibitem[\protect\citeauthoryear{{Galametz} et~al.,}{{Galametz}
  et~al.}{2018}]{Galametz2018}
{Galametz} M.,  et~al., 2018, \mn@doi [\aap] {10.1051/0004-6361/201833004},
  \href {http://adsabs.harvard.edu/abs/2018A%26A...616A.139G} {616, A139}

\bibitem[\protect\citeauthoryear{{Girart}, {Rao}  \& {Marrone}}{{Girart}
  et~al.}{2006}]{Girart2006}
{Girart} J.~M.,  {Rao} R.,   {Marrone} D.~P.,  2006, \mn@doi [Science]
  {10.1126/science.1129093}, \href
  {http://adsabs.harvard.edu/abs/2006Sci...313..812G} {313, 812}

\bibitem[\protect\citeauthoryear{{Gon{\c c}alves}, {Galli}  \&
  {Girart}}{{Gon{\c c}alves} et~al.}{2008}]{Goncalves2008}
{Gon{\c c}alves} J.,  {Galli} D.,   {Girart} J.~M.,  2008, \mn@doi [\aap]
  {10.1051/0004-6361:200810861}, \href
  {http://adsabs.harvard.edu/abs/2008A%26A...490L..39G} {490, L39}

\bibitem[\protect\citeauthoryear{Herbst}{Herbst}{2017}]{Herbst2017}
Herbst E.,  2017, \mn@doi [International Reviews in Physical Chemistry]
  {10.1080/0144235X.2017.1293974}, 36, 287

\bibitem[\protect\citeauthoryear{{Hildebrand} \& {Dragovan}}{{Hildebrand} \&
  {Dragovan}}{1995}]{Hildebrand95}
{Hildebrand} R.~H.,  {Dragovan} M.,  1995, \mn@doi [\apj] {10.1086/176173},
  \href {http://adsabs.harvard.edu/abs/1995ApJ...450..663H} {450, 663}

\bibitem[\protect\citeauthoryear{{Hoang} \& {Lazarian}}{{Hoang} \&
  {Lazarian}}{2014}]{Hoang14}
{Hoang} T.,  {Lazarian} A.,  2014, \mn@doi [\mnras] {10.1093/mnras/stt2240},
  \href {http://adsabs.harvard.edu/abs/2014MNRAS.438..680H} {438, 680}

\bibitem[\protect\citeauthoryear{{Hoang} \& {Lazarian}}{{Hoang} \&
  {Lazarian}}{2016}]{Hoang2016}
{Hoang} T.,  {Lazarian} A.,  2016, \mn@doi [\apj]
  {10.3847/0004-637X/831/2/159}, \href
  {http://adsabs.harvard.edu/abs/2016ApJ...831..159H} {831, 159}

\bibitem[\protect\citeauthoryear{{Hull} et~al.,}{{Hull}
  et~al.}{2014}]{Hull2014}
{Hull} C.~L.~H.,  et~al., 2014, \mn@doi [\apjs] {10.1088/0067-0049/213/1/13},
  \href {http://adsabs.harvard.edu/abs/2014ApJS..213...13H} {213, 13}

\bibitem[\protect\citeauthoryear{{Hull} et~al.,}{{Hull}
  et~al.}{2017a}]{Hull2017}
{Hull} C. L.~H.,  et~al., 2017a, \mn@doi [\apj] {10.3847/1538-4357/aa7fe9},
  \href {https://ui.adsabs.harvard.edu/\#abs/2017ApJ...847...92H} {847, 92}

\bibitem[\protect\citeauthoryear{{Hull} et~al.,}{{Hull}
  et~al.}{2017b}]{Hull2017b}
{Hull} C.~L.~H.,  et~al., 2017b, \mn@doi [\apj] {10.3847/1538-4357/aa7fe9},
  \href {http://adsabs.harvard.edu/abs/2017ApJ...847...92H} {847, 92}

\bibitem[\protect\citeauthoryear{{Jones}, {Fanciullo}, {K{\"o}hler},
  {Verstraete}, {Guillet}, {Bocchio}  \& {Ysard}}{{Jones}
  et~al.}{2013}]{Jones2013}
{Jones} A.~P.,  {Fanciullo} L.,  {K{\"o}hler} M.,  {Verstraete} L.,  {Guillet}
  V.,  {Bocchio} M.,   {Ysard} N.,  2013, \mn@doi [\aap]
  {10.1051/0004-6361/201321686}, \href
  {https://ui.adsabs.harvard.edu/\#abs/2013A&A...558A..62J} {558, A62}

\bibitem[\protect\citeauthoryear{{Kataoka} et~al.,}{{Kataoka}
  et~al.}{2015}]{Kataoka2015}
{Kataoka} A.,  et~al., 2015, \mn@doi [\apj] {10.1088/0004-637X/809/1/78}, \href
  {https://ui.adsabs.harvard.edu/abs/2015ApJ...809...78K} {809, 78}

\bibitem[\protect\citeauthoryear{{K{\"o}nyves} et~al.,}{{K{\"o}nyves}
  et~al.}{2015}]{Konyves2015}
{K{\"o}nyves} V.,  et~al., 2015, \mn@doi [\aap] {10.1051/0004-6361/201525861},
  \href {https://ui.adsabs.harvard.edu/\#abs/2015A&A...584A..91K} {584, A91}

\bibitem[\protect\citeauthoryear{{Kwon}, {Looney}, {Mundy}, {Chiang}  \&
  {Kemball}}{{Kwon} et~al.}{2009}]{Kwon2009}
{Kwon} W.,  {Looney} L.~W.,  {Mundy} L.~G.,  {Chiang} H.-F.,   {Kemball} A.~J.,
   2009, \mn@doi [\apj] {10.1088/0004-637X/696/1/841}, \href
  {https://ui.adsabs.harvard.edu/abs/2009ApJ...696..841K} {696, 841}

\bibitem[\protect\citeauthoryear{{Lazarian}}{{Lazarian}}{2009}]{Lazarian2009}
{Lazarian} A.,  2009, in {Henning} T.,  {Gr{\"u}n} E.,   {Steinacker} J.,  eds,
   ASP Conf. Series Vol. 414, Cosmic Dust - Near and Far. p.~482 (\mn@eprint
  {arXiv} {0903.1100})

\bibitem[\protect\citeauthoryear{{Lazarian} \& {Hoang}}{{Lazarian} \&
  {Hoang}}{2007}]{LazarianHoang07}
{Lazarian} A.,  {Hoang} T.,  2007, \mn@doi [\mnras]
  {10.1111/j.1365-2966.2007.11817.x}, \href
  {http://adsabs.harvard.edu/abs/2007MNRAS.378..910L} {378, 910}

\bibitem[\protect\citeauthoryear{{Lef{\`e}vre}, {Pagani}, {Min}, {Poteet}  \&
  {Whittet}}{{Lef{\`e}vre} et~al.}{2016}]{Lefevre2016}
{Lef{\`e}vre} C.,  {Pagani} L.,  {Min} M.,  {Poteet} C.,   {Whittet} D.,  2016,
  \mn@doi [\aap] {10.1051/0004-6361/201526999}, \href
  {https://ui.adsabs.harvard.edu/\#abs/2016A&A...585L...4L} {585, L4}

\bibitem[\protect\citeauthoryear{{Martin} et~al.,}{{Martin}
  et~al.}{2012}]{Martin2012}
{Martin} P.~G.,  et~al., 2012, \mn@doi [\apj] {10.1088/0004-637X/751/1/28},
  \href {https://ui.adsabs.harvard.edu/\#abs/2012ApJ...751...28M} {751, 28}

\bibitem[\protect\citeauthoryear{{Mathis}, {Rumpl}  \& {Nordsieck}}{{Mathis}
  et~al.}{1977}]{Mathis1977}
{Mathis} J.~S.,  {Rumpl} W.,   {Nordsieck} K.~H.,  1977, \mn@doi [\apj]
  {10.1086/155591}, \href {http://adsabs.harvard.edu/abs/1977ApJ...217..425M}
  {217, 425}

\bibitem[\protect\citeauthoryear{{Mathis}, {Mezger}  \& {Panagia}}{{Mathis}
  et~al.}{1983}]{Mathis1983}
{Mathis} J.~S.,  {Mezger} P.~G.,   {Panagia} N.,  1983, \aap, \href
  {http://adsabs.harvard.edu/abs/1983A%26A...128..212M} {128, 212}

\bibitem[\protect\citeauthoryear{{Maury} et~al.,}{{Maury}
  et~al.}{2018}]{Maury2018}
{Maury} A.~J.,  et~al., 2018, \mn@doi [\mnras] {10.1093/mnras/sty574}, \href
  {http://adsabs.harvard.edu/abs/2018MNRAS.477.2760M} {477, 2760}

\bibitem[\protect\citeauthoryear{{Maury} et~al.,}{{Maury}
  et~al.}{2019}]{Maury2019}
{Maury} A.~J.,  et~al., 2019, \mn@doi [\aap] {10.1051/0004-6361/201833537},
  \href {http://cdsads.u-strasbg.fr/abs/2019A%26A...621A..76M} {621, A76}

\bibitem[\protect\citeauthoryear{{Min}, {Dullemond}, {Dominik}, {de Koter}  \&
  {Hovenier}}{{Min} et~al.}{2009}]{Min2009}
{Min} M.,  {Dullemond} C.~P.,  {Dominik} C.,  {de Koter} A.,   {Hovenier}
  J.~W.,  2009, \mn@doi [\aap] {10.1051/0004-6361/200811470}, \href
  {http://adsabs.harvard.edu/abs/2009A%26A...497..155M} {497, 155}

\bibitem[\protect\citeauthoryear{{Ormel}, {Min}, {Tielens}, {Dominik}  \&
  {Paszun}}{{Ormel} et~al.}{2011}]{Ormel2011}
{Ormel} C.~W.,  {Min} M.,  {Tielens} A.~G.~G.~M.,  {Dominik} C.,   {Paszun} D.,
   2011, \mn@doi [\aap] {10.1051/0004-6361/201117058}, \href
  {http://adsabs.harvard.edu/abs/2011A%26A...532A..43O} {532, A43}

\bibitem[\protect\citeauthoryear{{Padovani} et~al.,}{{Padovani}
  et~al.}{2012}]{Padovani2012}
{Padovani} M.,  et~al., 2012, \mn@doi [\aap] {10.1051/0004-6361/201219028},
  \href {https://ui.adsabs.harvard.edu/\#abs/2012A&A...543A..16P} {543, A16}

\bibitem[\protect\citeauthoryear{{Pagani}, {Steinacker}, {Bacmann}, {Stutz}  \&
  {Henning}}{{Pagani} et~al.}{2010}]{Pagani2010}
{Pagani} L.,  {Steinacker} J.,  {Bacmann} A.,  {Stutz} A.,   {Henning} T.,
  2010, \mn@doi [Science] {10.1126/science.1193211}, \href
  {http://adsabs.harvard.edu/abs/2010Sci...329.1622P} {329, 1622}

\bibitem[\protect\citeauthoryear{{Planck Collaboration} et~al.,}{{Planck
  Collaboration} et~al.}{2015}]{PlanckXXI2015}
{Planck Collaboration} et~al., 2015, \mn@doi [\aap]
  {10.1051/0004-6361/201424087}, \href
  {https://ui.adsabs.harvard.edu/\#abs/2015A&A...576A.106P} {576, A106}

\bibitem[\protect\citeauthoryear{{Purcell}}{{Purcell}}{1979}]{Purcell79}
{Purcell} E.~M.,  1979, \mn@doi [\apj] {10.1086/157204}, \href
  {http://adsabs.harvard.edu/abs/1979ApJ...231..404P} {231, 404}

\bibitem[\protect\citeauthoryear{{Rao}, {Girart}, {Marrone}, {Lai}  \&
  {Schnee}}{{Rao} et~al.}{2009}]{Rao2009}
{Rao} R.,  {Girart} J.~M.,  {Marrone} D.~P.,  {Lai} S.-P.,   {Schnee} S.,
  2009, \mn@doi [\apj] {10.1088/0004-637X/707/2/921}, \href
  {http://adsabs.harvard.edu/abs/2009ApJ...707..921R} {707, 921}

\bibitem[\protect\citeauthoryear{{Reissl}, {Wolf}  \& {Brauer}}{{Reissl}
  et~al.}{2016}]{Reissl16}
{Reissl} S.,  {Wolf} S.,   {Brauer} R.,  2016, \mn@doi [\aap]
  {10.1051/0004-6361/201424930}, \href
  {http://adsabs.harvard.edu/abs/2016A%26A...593A..87R} {593, A87}

\bibitem[\protect\citeauthoryear{{Roberge}}{{Roberge}}{2004}]{Roberge2004}
{Roberge} W.~G.,  2004, in {Witt} A.~N.,  {Clayton} G.~C.,   {Draine} B.~T.,
  eds,  ASP Conf. Series Vol. 309, Astrophysics of Dust. p.~467

\bibitem[\protect\citeauthoryear{{Sadavoy} et~al.,}{{Sadavoy}
  et~al.}{2018}]{Sadavoy2018}
{Sadavoy} S.~I.,  et~al., 2018, \mn@doi [\apj] {10.3847/1538-4357/aac21a},
  \href {http://adsabs.harvard.edu/abs/2018ApJ...859..165S} {859, 165}

\bibitem[\protect\citeauthoryear{{Segura-Cox}, {Looney}, {Stephens},
  {Fern{\'a}ndez-L{\'o}pez}, {Kwon}, {Tobin}, {Li}  \& {Crutcher}}{{Segura-Cox}
  et~al.}{2015}]{Segura-Cox2015}
{Segura-Cox} D.~M.,  {Looney} L.~W.,  {Stephens} I.~W.,
  {Fern{\'a}ndez-L{\'o}pez} M.,  {Kwon} W.,  {Tobin} J.~J.,  {Li} Z.-Y.,
  {Crutcher} R.,  2015, \mn@doi [\apjl] {10.1088/2041-8205/798/1/L2}, \href
  {http://adsabs.harvard.edu/abs/2015ApJ...798L...2S} {798, L2}

\bibitem[\protect\citeauthoryear{{Soam} et~al.,}{{Soam}
  et~al.}{2018}]{Soam2018}
{Soam} A.,  et~al., 2018, \mn@doi [\apj] {10.3847/1538-4357/aac4a6}, \href
  {http://adsabs.harvard.edu/abs/2018ApJ...861...65S} {861, 65}

\bibitem[\protect\citeauthoryear{{Steinacker}, {Pagani}, {Bacmann}  \&
  {Guieu}}{{Steinacker} et~al.}{2010}]{Steinacker2010}
{Steinacker} J.,  {Pagani} L.,  {Bacmann} A.,   {Guieu} S.,  2010, \mn@doi
  [\aap] {10.1051/0004-6361/200912835}, \href
  {https://ui.adsabs.harvard.edu/\#abs/2010A&A...511A...9S} {511, A9}

\bibitem[\protect\citeauthoryear{{Stephens} et~al.,}{{Stephens}
  et~al.}{2013}]{Stephens2013}
{Stephens} I.~W.,  et~al., 2013, \mn@doi [\apjl] {10.1088/2041-8205/769/1/L15},
  \href {http://adsabs.harvard.edu/abs/2013ApJ...769L..15S} {769, L15}

\bibitem[\protect\citeauthoryear{{Stepnik} et~al.,}{{Stepnik}
  et~al.}{2003}]{Stepnik2003}
{Stepnik} B.,  et~al., 2003, \mn@doi [\aap] {10.1051/0004-6361:20021309}, \href
  {http://adsabs.harvard.edu/abs/2003A%26A...398..551S} {398, 551}

\bibitem[\protect\citeauthoryear{{Testi} et~al.,}{{Testi}
  et~al.}{2014}]{Testi2014}
{Testi} L.,  et~al., 2014, \mn@doi [Protostars and Planets VI]
  {10.2458/azu_uapress_9780816531240-ch015}, \href
  {https://ui.adsabs.harvard.edu/\#abs/2014prpl.conf..339T} {p.~339}

\bibitem[\protect\citeauthoryear{{Teyssier}}{{Teyssier}}{2002}]{Teyssier2002}
{Teyssier} R.,  2002, \mn@doi [\aap] {10.1051/0004-6361:20011817}, \href
  {http://adsabs.harvard.edu/abs/2002A%26A...385..337T} {385, 337}

\bibitem[\protect\citeauthoryear{{Wolf}, {Launhardt}  \& {Henning}}{{Wolf}
  et~al.}{2003}]{Wolf2003}
{Wolf} S.,  {Launhardt} R.,   {Henning} T.,  2003, \mn@doi [\apj]
  {10.1086/375622}, \href {http://adsabs.harvard.edu/abs/2003ApJ...592..233W}
  {592, 233}

\bibitem[\protect\citeauthoryear{Wurster \& Li}{Wurster \&
  Li}{2018}]{Wurster2018}
Wurster J.,  Li Z.-Y.,  2018, \mn@doi [Frontiers in Astronomy and Space
  Sciences] {10.3389/fspas.2018.00039}, 5, 39

\bibitem[\protect\citeauthoryear{{Yen} et~al.,}{{Yen} et~al.}{2019}]{Yen2019}
{Yen} H.-W.,  et~al., 2019, \mn@doi [\apj] {10.3847/1538-4357/aafb6c}, \href
  {http://adsabs.harvard.edu/abs/2019ApJ...871..243Y} {871, 243}

\bibitem[\protect\citeauthoryear{{Ysard}, {K{\"o}hler}, {Jones}, {Dartois},
  {Godard}  \& {Gavilan}}{{Ysard} et~al.}{2016}]{Ysard2016}
{Ysard} N.,  {K{\"o}hler} M.,  {Jones} A.,  {Dartois} E.,  {Godard} M.,
  {Gavilan} L.,  2016, \mn@doi [\aap] {10.1051/0004-6361/201527487}, \href
  {http://adsabs.harvard.edu/abs/2016A%26A...588A..44Y} {588, A44}

\bibitem[\protect\citeauthoryear{{Zhao}, {Caselli}  \& {Li}}{{Zhao}
  et~al.}{2018}]{Zhao2018}
{Zhao} B.,  {Caselli} P.,   {Li} Z.-Y.,  2018, \mn@doi [\mnras]
  {10.1093/mnras/sty1165}, \href
  {http://cdsads.u-strasbg.fr/abs/2018MNRAS.478.2723Z} {478, 2723}

\makeatother
\end{thebibliography}

%\begin{thebibliography}{99}
%\end{thebibliography}

%%%%%%%%%%%%%%%%%%%%%%%%%%%%%%%%%%%%%%%%%%%%%%%%%%

%%%%%%%%%%%%%%%%% APPENDICES %%%%%%%%%%%%%%%%%%%%%

%\iffalse
\appendix

\section{Some extra material}

\begin{figure}
\centering
\includegraphics[width=0.45\textwidth, trim={4cm 1.5cm 6cm 1.6cm},clip]{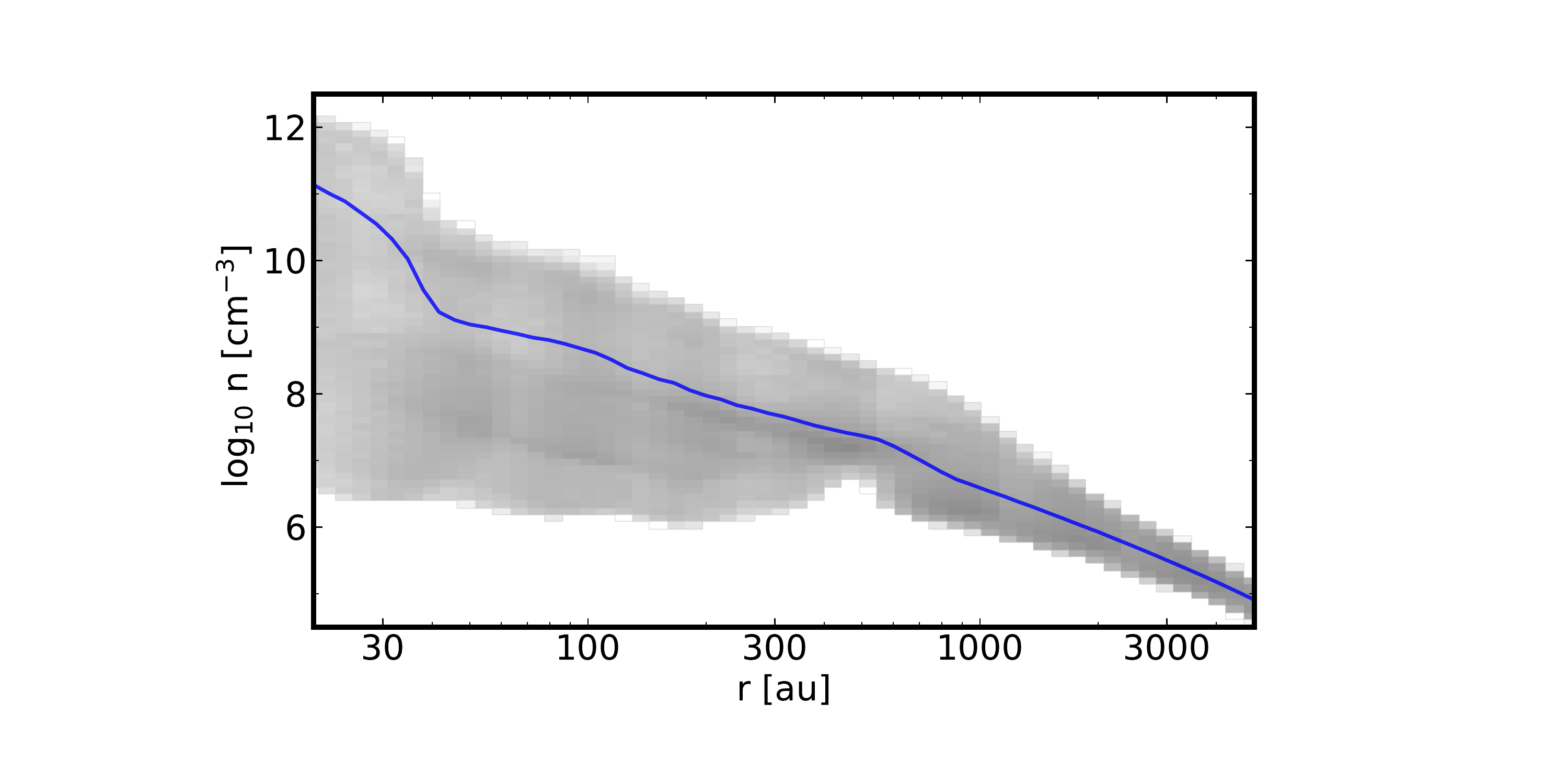}
\caption{Local total gas number density radial distribution from the non-ideal MHD simulation.
The radial number density profile shows the mean value (blue line) and the dispersion (gray) for the selected output. %\textcolor{green}{serait bien de mettre une double echelle dans la Figure A1 et montrer la courbe radiale de densit\'e avec les variations locales comme une dispersion en gris\'e ?}
}
\label{profile_dens}
\end{figure}

\begin{figure}
\centering
\begin{tikzpicture}
\node[above right] (img) at (0,0) {
  % trim left, bottom, right, top
  \includegraphics[width=0.4\textwidth, trim={0.75cm 0.5cm 1.cm 1.2cm},clip]{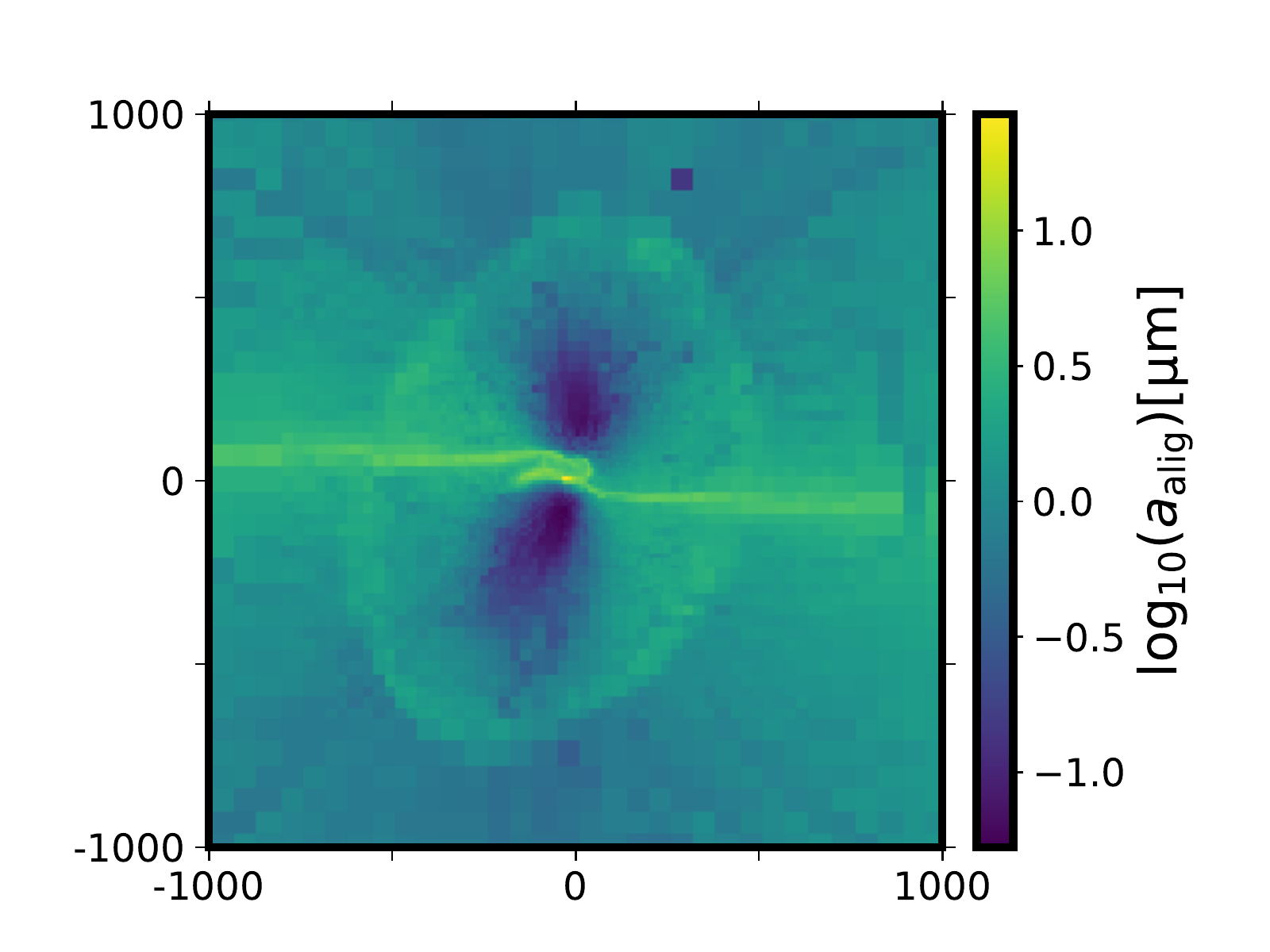}};
  \node [rotate=90, scale=0.8] at (0.3,2.8) {\fontfamily{phv}\selectfont [au]};
    \node [rotate=0, scale=0.8] at (3.35,0.05) {\fontfamily{phv}\selectfont [au]};
\end{tikzpicture}
\caption{Mean $a_\mathrm{alig}$ in the central $50\au$ region for the $a_\mathrm{max}=30~\micron$ distribution. 
$a_\mathrm{alig}$ is a local value and it corresponds to the minimum size at which dust grains start to align according to the local conditions in the RATs theory. In the cavities created by the outflow grains as small as $0.1~\micron$ can start to align, while in the equatorial region, only grains larger than $10~\micron$ can be aligned.
The value shown in this figure corresponds to the mean computed in a slab (perpendicular to the line-of-sight) of thickness $50\au$ around the center of the simulation box.  
}
\label{aalig}
\end{figure}

%\subsection{Dust temperature}
\begin{figure}
\centering
\includegraphics[width=0.45\textwidth, trim={4cm 1.5cm 6cm 1.6cm},clip]{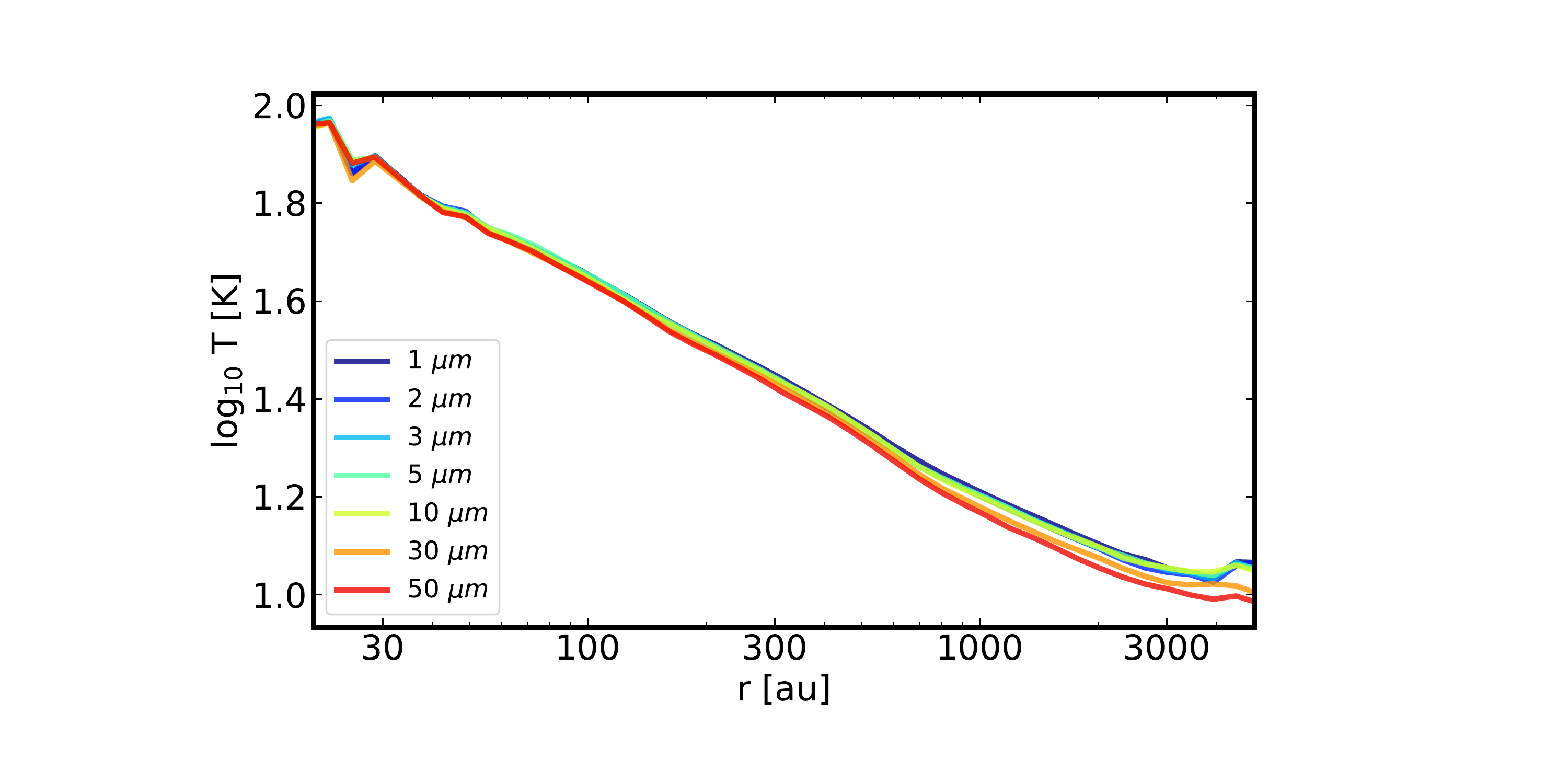}
\caption{Radial mean dust temperature profiles for different distributions. $a_\mathrm{max}$ is given in the figure. The dust temperature do not strongly depend on the dust size distribution. 
%\textcolor{green}{serait bien de mettre une double echelle dans la Figure A1 et montrer la courbe radiale de densit\'e avec les variations locales comme une dispersion en gris\'e ?}
}
\label{profile_temp}
\end{figure}

\begin{figure}
\centering
\begin{tabular}{@{}l@{}}
\includegraphics[width=0.45\textwidth, trim={4cm 2.2cm 6cm 1.6cm},clip]{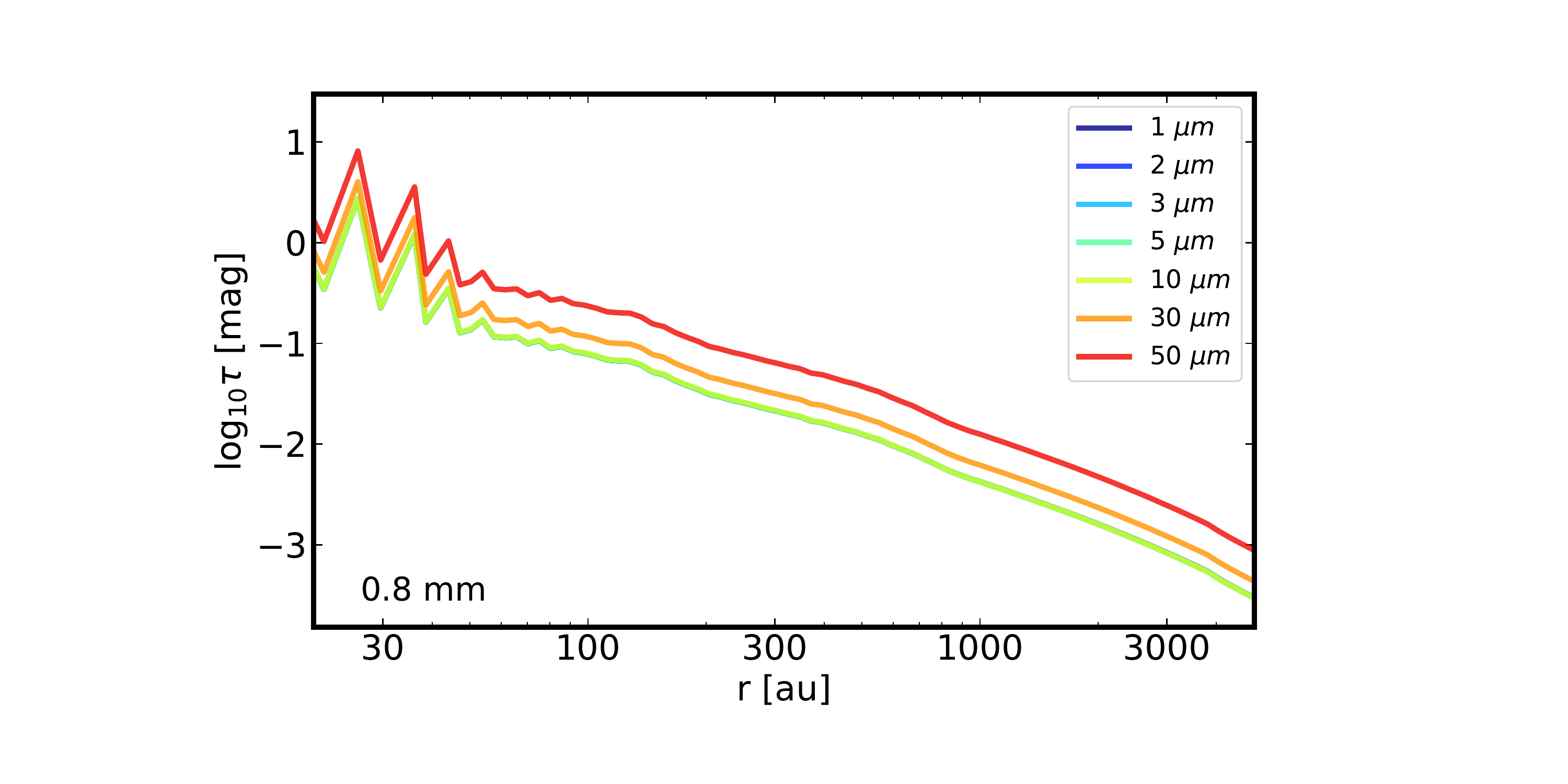}\\
\includegraphics[width=0.45\textwidth, trim={4cm 1.5cm 6cm 1.6cm},clip]{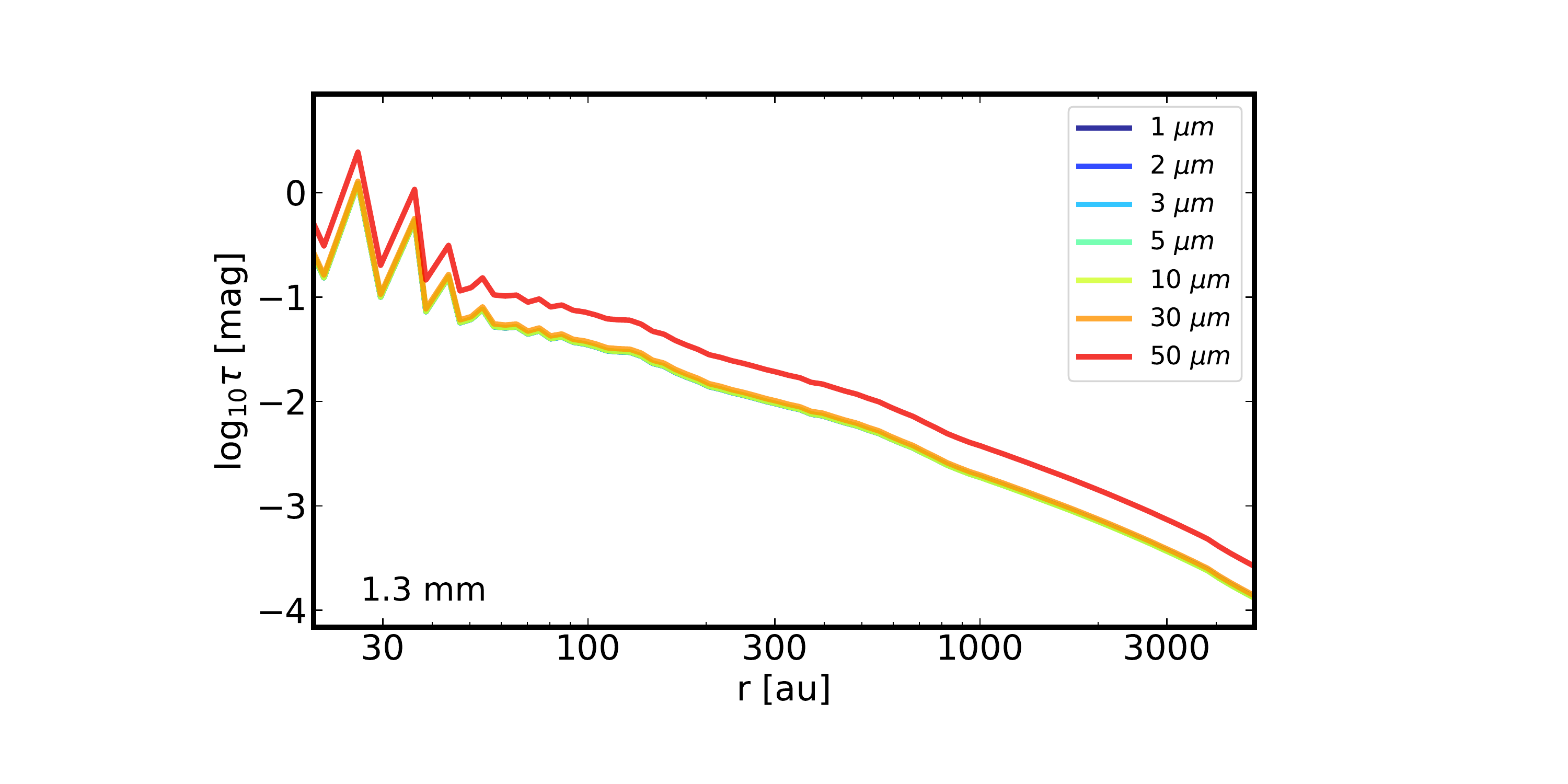}
\end{tabular}
\caption{Radial mean opacity profiles at two wavelengths: $\lambda = 0.8$ (top), and $1.3~\mathrm{mm}$ for all the dust distributions used in this paper. $a_\mathrm{max}$ is given in the figure. }
\label{profile_opacity}
\end{figure}

%\subsection{Synthetic observations}
\begin{figure*}
\centering
\hspace{-0.7cm}
\begin{tikzpicture}
\node[above right] (img) at (0,0) {
  \begin{tabular}{@{}llll@{}}
  % trim left, bottom, right, top
  \includegraphics[height=0.22\textwidth, trim={0.75cm 1.1cm 4.05cm 1.2cm},clip]{Figures/RAT_2Lsun_big2_00051_pix200_log_poliI2.pdf}&
  \includegraphics[height=0.22\textwidth, trim={2.5cm 1.1cm 4.05cm 1.2cm},clip]{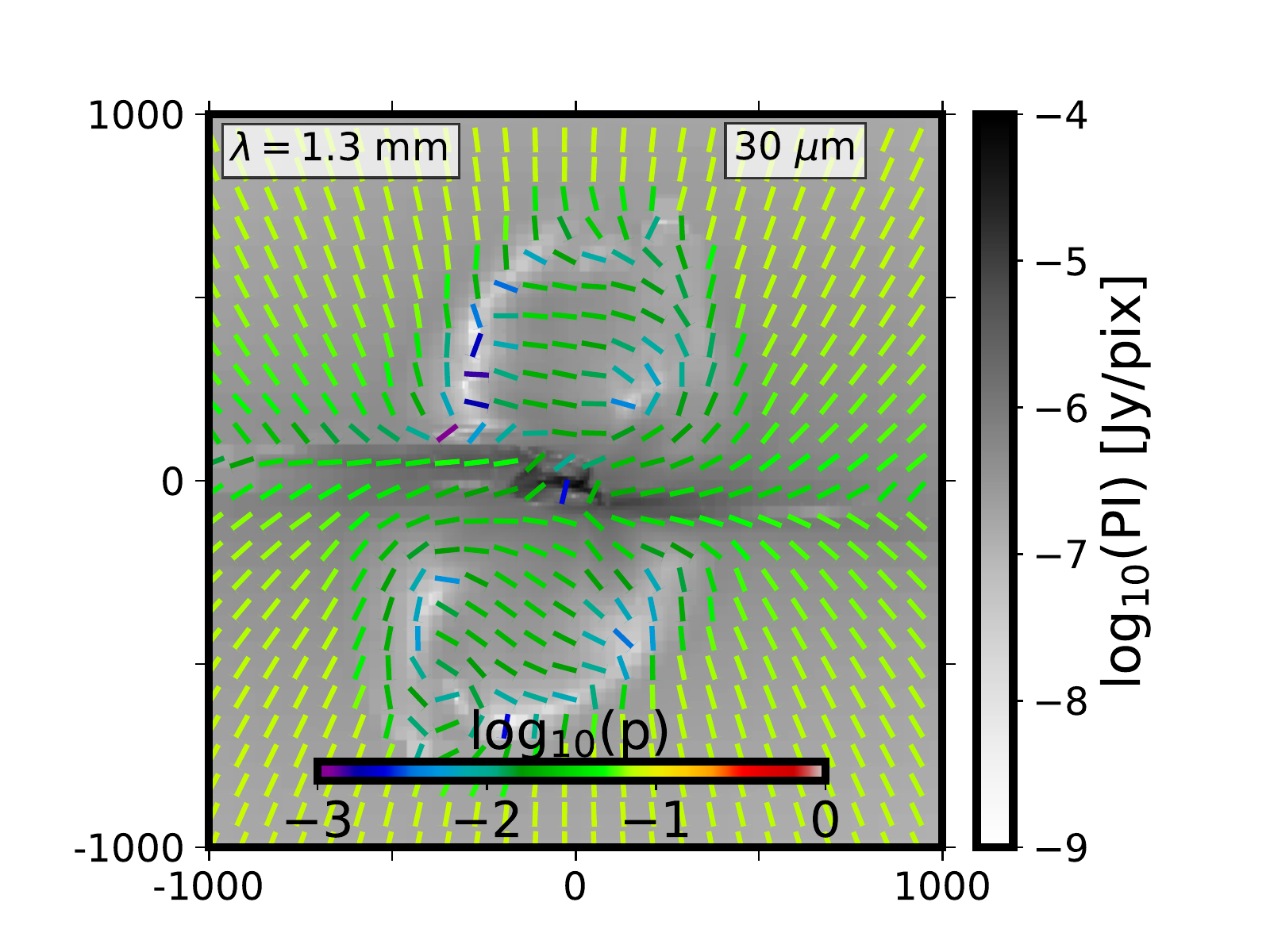}&
  \includegraphics[height=0.22\textwidth, trim={2.5cm 1.1cm 4.05cm 1.2cm},clip]{Figures/RAT_2Lsun_big3_00051_pix200_log_poliI2.pdf}&
  \includegraphics[height=0.22\textwidth, trim={2.5cm 1.1cm 4.05cm 1.2cm},clip]{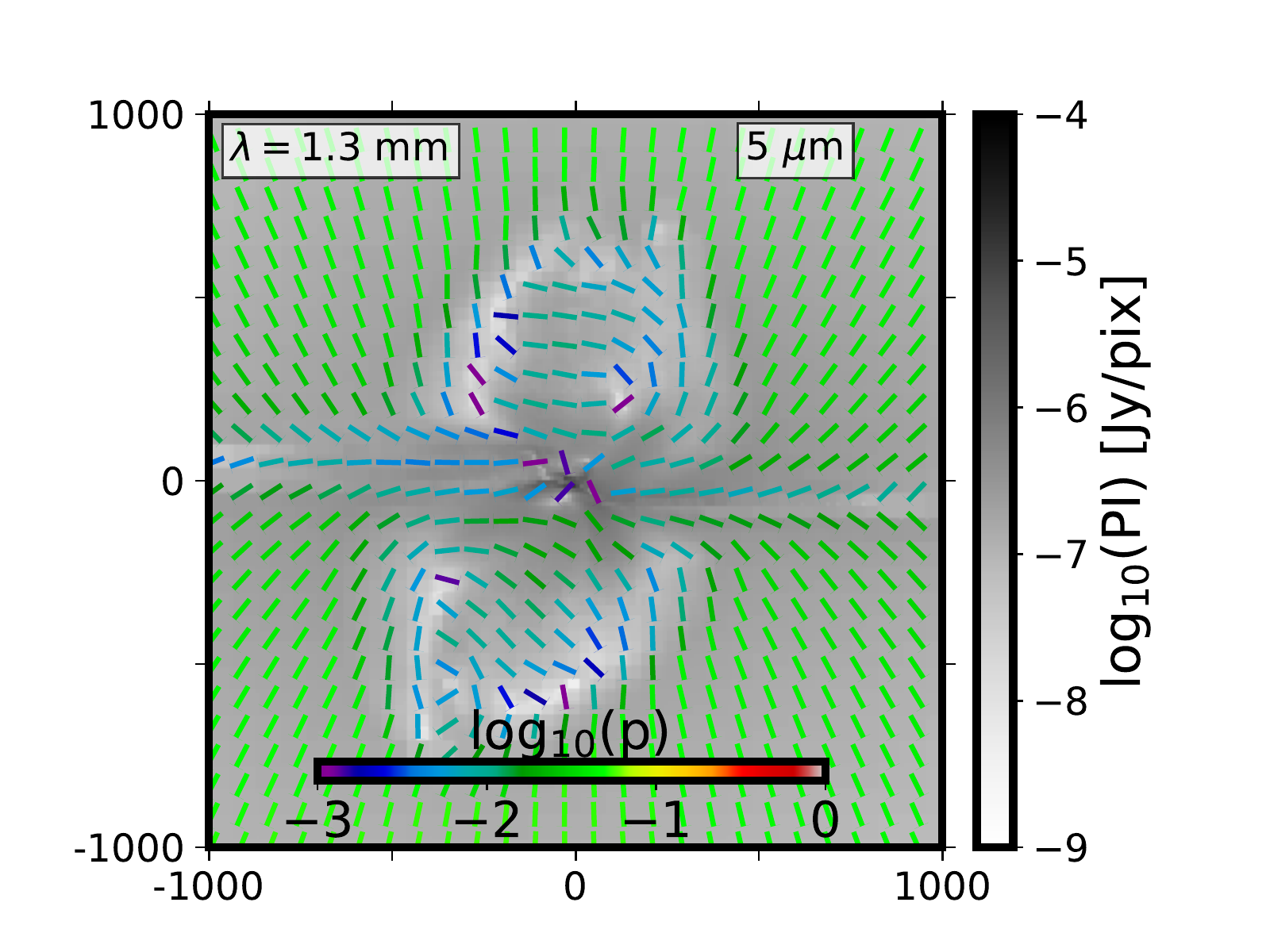} \\
  \includegraphics[height=0.22\textwidth, trim={0.75cm 1.1cm 4.05cm 1.2cm},clip]{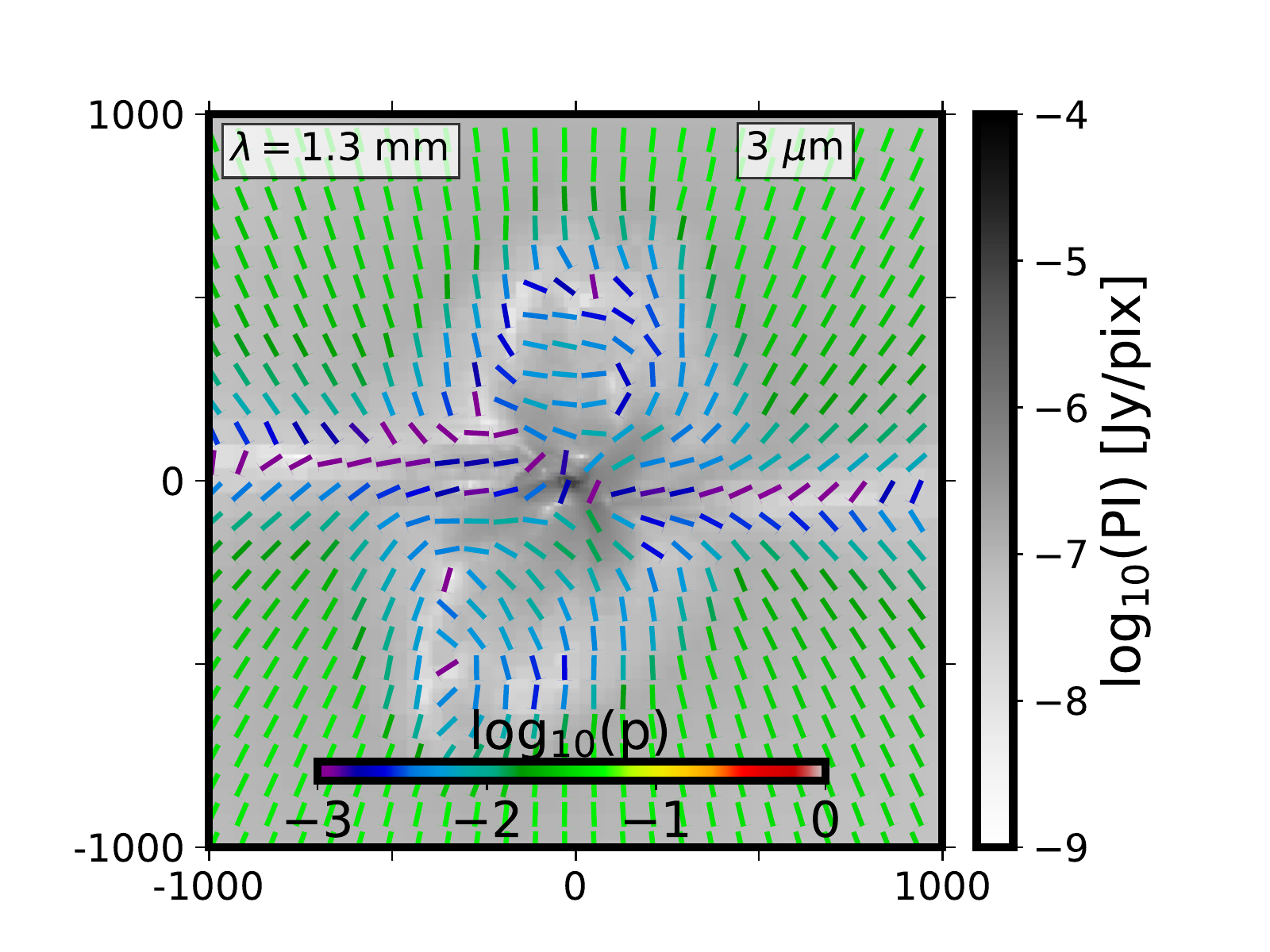}&
  \includegraphics[height=0.22\textwidth, trim={2.5cm 1.1cm 4.05cm 1.2cm},clip]{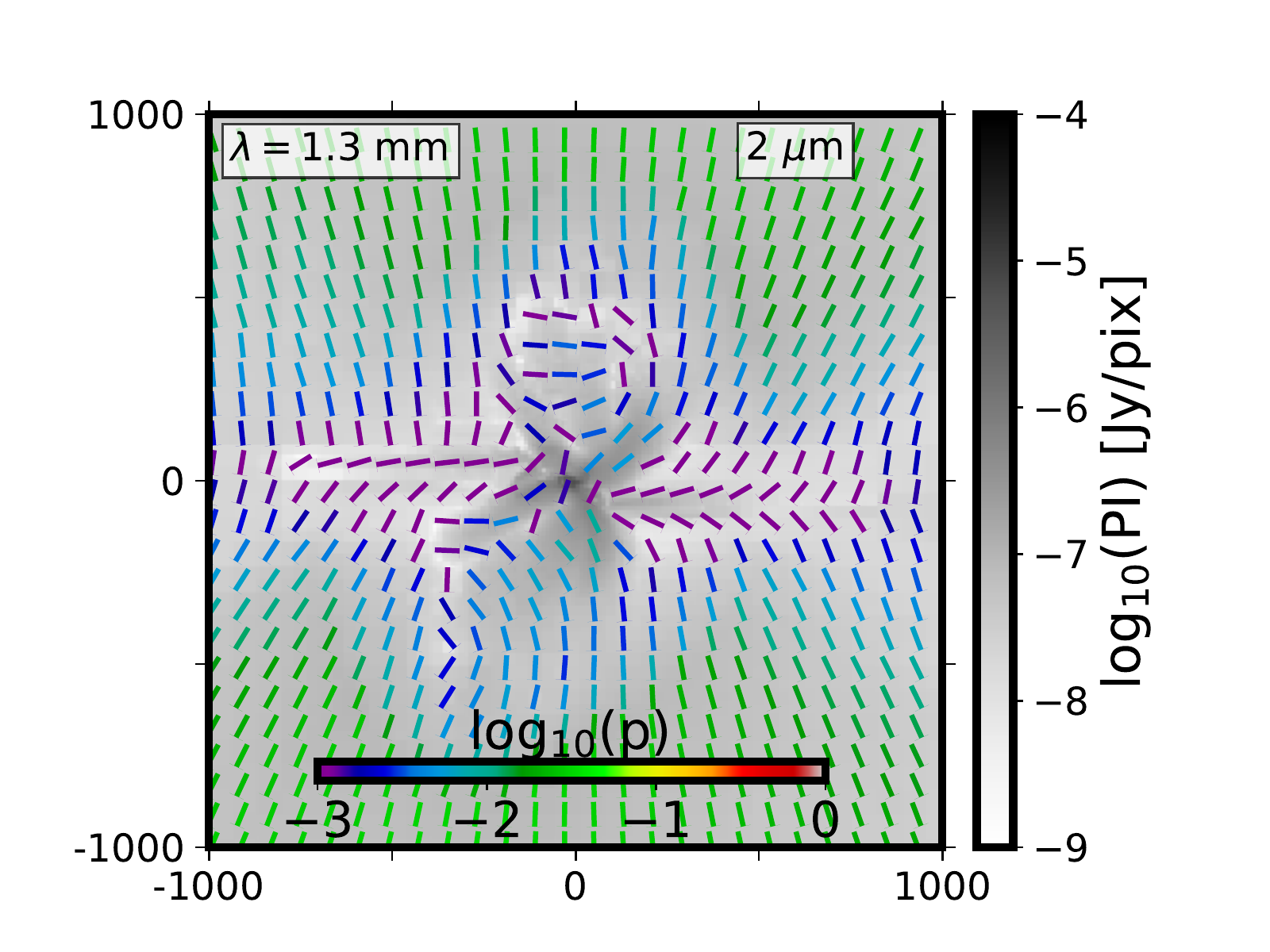}&
  \includegraphics[height=0.22\textwidth, trim={2.5cm 1.1cm 4.05cm 1.2cm},clip]{Figures/RAT_2Lsun_big7_00051_pix200_log_poliI2.pdf}&
  \includegraphics[height=0.22\textwidth, trim={12.4cm 1.1cm 0cm 1.2cm},clip]{Figures/RAT_2Lsun_big7_00051_pix200_log_poliI2.pdf}  
  \end{tabular}};
  \node [rotate=90, scale=0.8] at (0.7,2.25) {\fontfamily{phv}\selectfont [au]};
  \node [rotate=90, scale=0.8] at (0.7,6.2) {\fontfamily{phv}\selectfont [au]};
    %\node [rotate=0, scale=0.8] at (3.35,0.05) {\fontfamily{phv}\selectfont [au]};
\end{tikzpicture}
\caption{Full set of perfect synthetic observations at $\lambda=1.3~\mathrm{mm}$. The polarized dust emission ($PI$) is shown in gray-scale as a background image, while the inferred magnetic field orientation and polarization fraction ($p$) are shown as overlaid color coded segments. $a_\mathrm{max}$ is indicated at the top-right corner of each panel. }
\label{Full_synth}
\end{figure*}

\begin{figure}
\centering
  \begin{tabular}{@{}l@{}}
% trim left, bottom, right, top
    \includegraphics[width=0.4\textwidth, trim={3.5cm 2.3cm 6.05cm 1.6cm},clip]{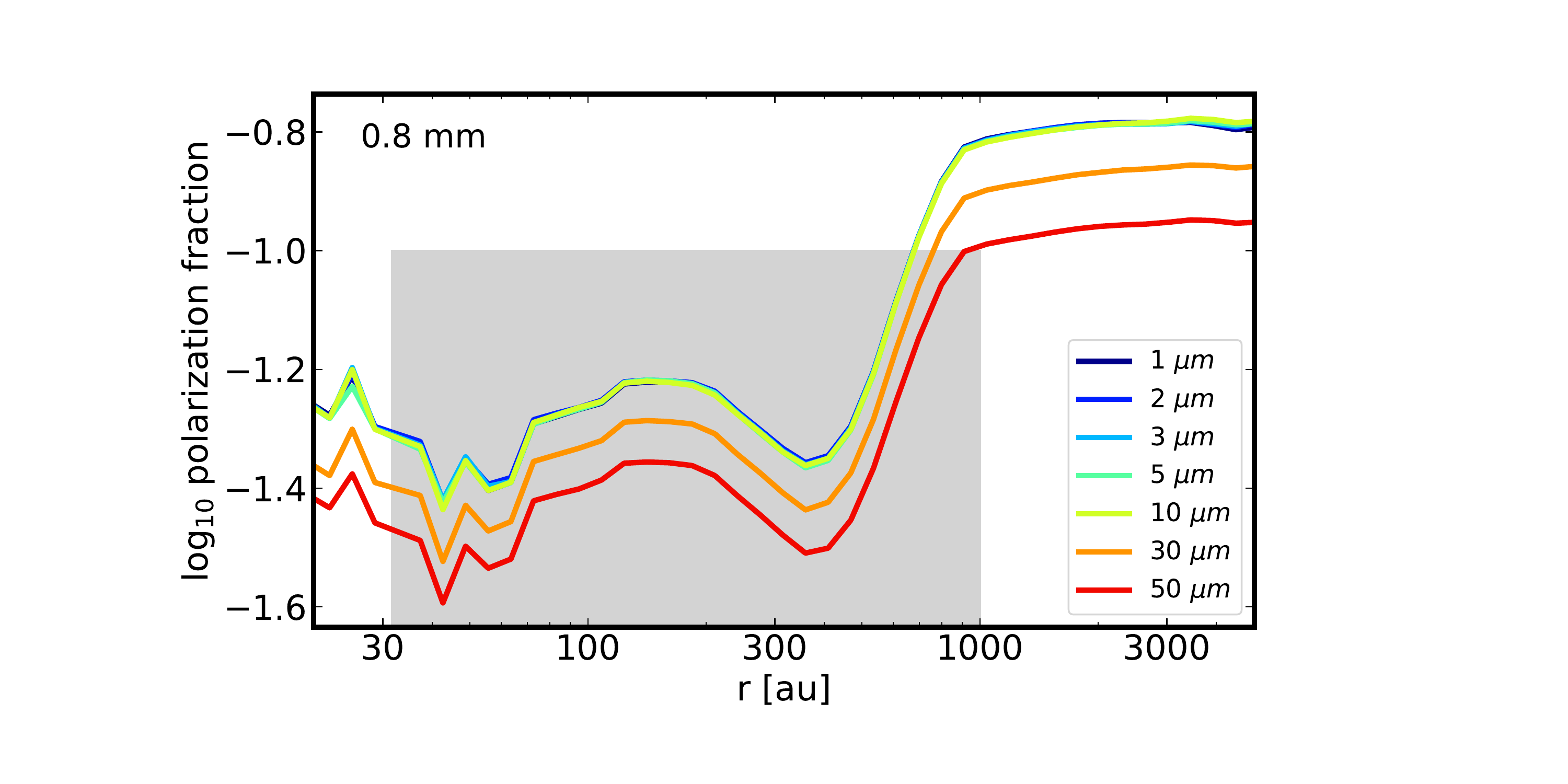}\\
    \includegraphics[width=0.4\textwidth, trim={3.5cm 1.2cm 6.05cm 1.6cm},clip]{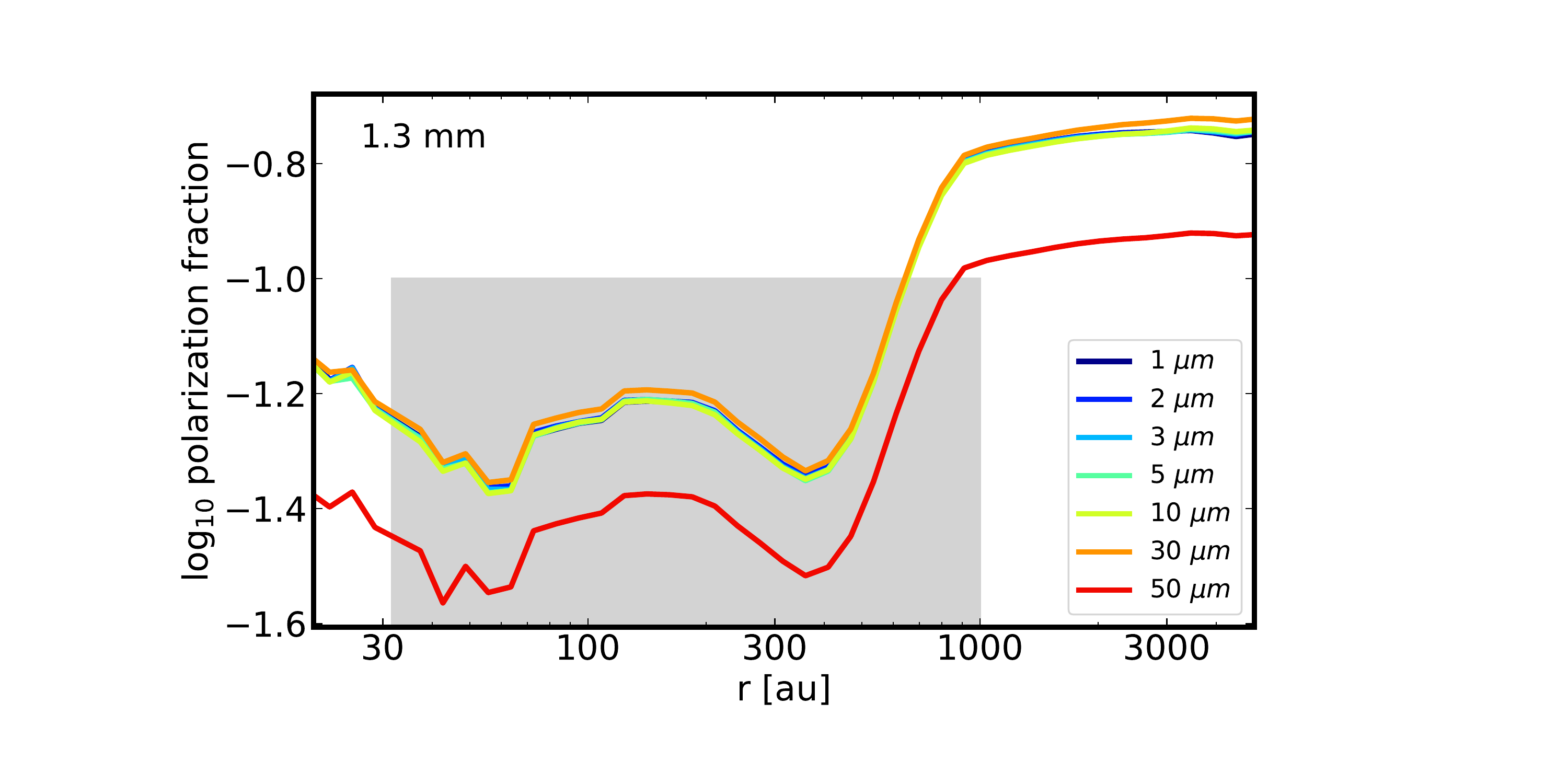}
  \end{tabular}
\caption{Radial mean polarization fraction assuming that all dust grains able to align (silicates) are perfectly aligned with the magnetic field. Results are shown for two wavelengths: $\lambda = 0.8~\mathrm{mm}$ (top panel), and $\lambda = 1.3~\mathrm{mm}$ (bottom panel). The maximum dust size for each distribution is given in the figure. The typical values observed in YSOs are shown as a grey region. %\textcolor{red}{SAY FOR WHICH DUST GRAIN SIZE DISTRIBUTION THIS WAS DONE} 
}
\label{radialPA}
\end{figure}

\begin{figure}
\centering
  \begin{tabular}{@{}l@{}}
% trim left, bottom, right, top
    \includegraphics[width=0.4\textwidth, trim={3.1cm 2.3cm 6.05cm 1.6cm},clip]{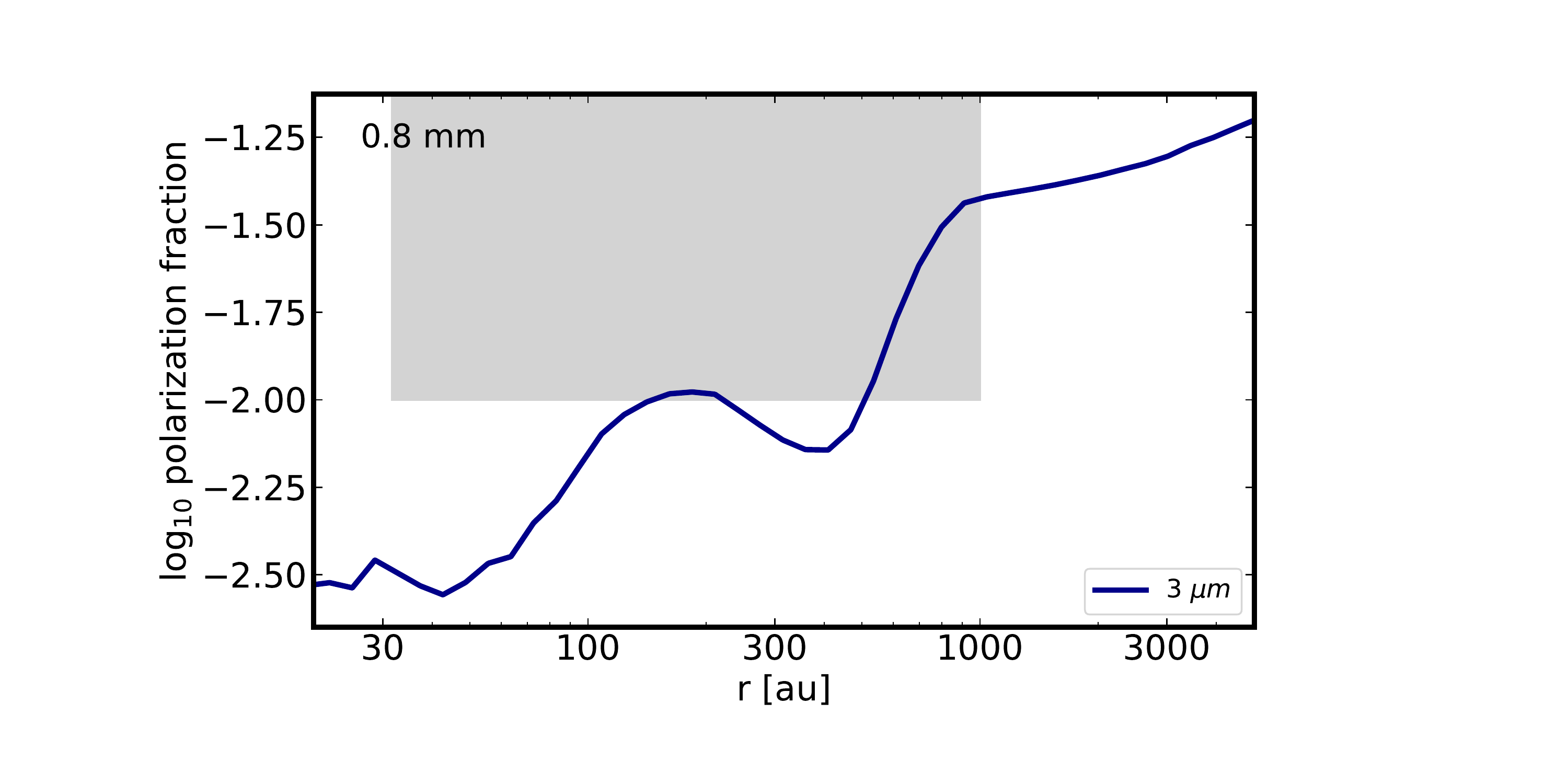}\\
    \includegraphics[width=0.4\textwidth, trim={3.1cm 1.2cm 6.05cm 1.6cm},clip]{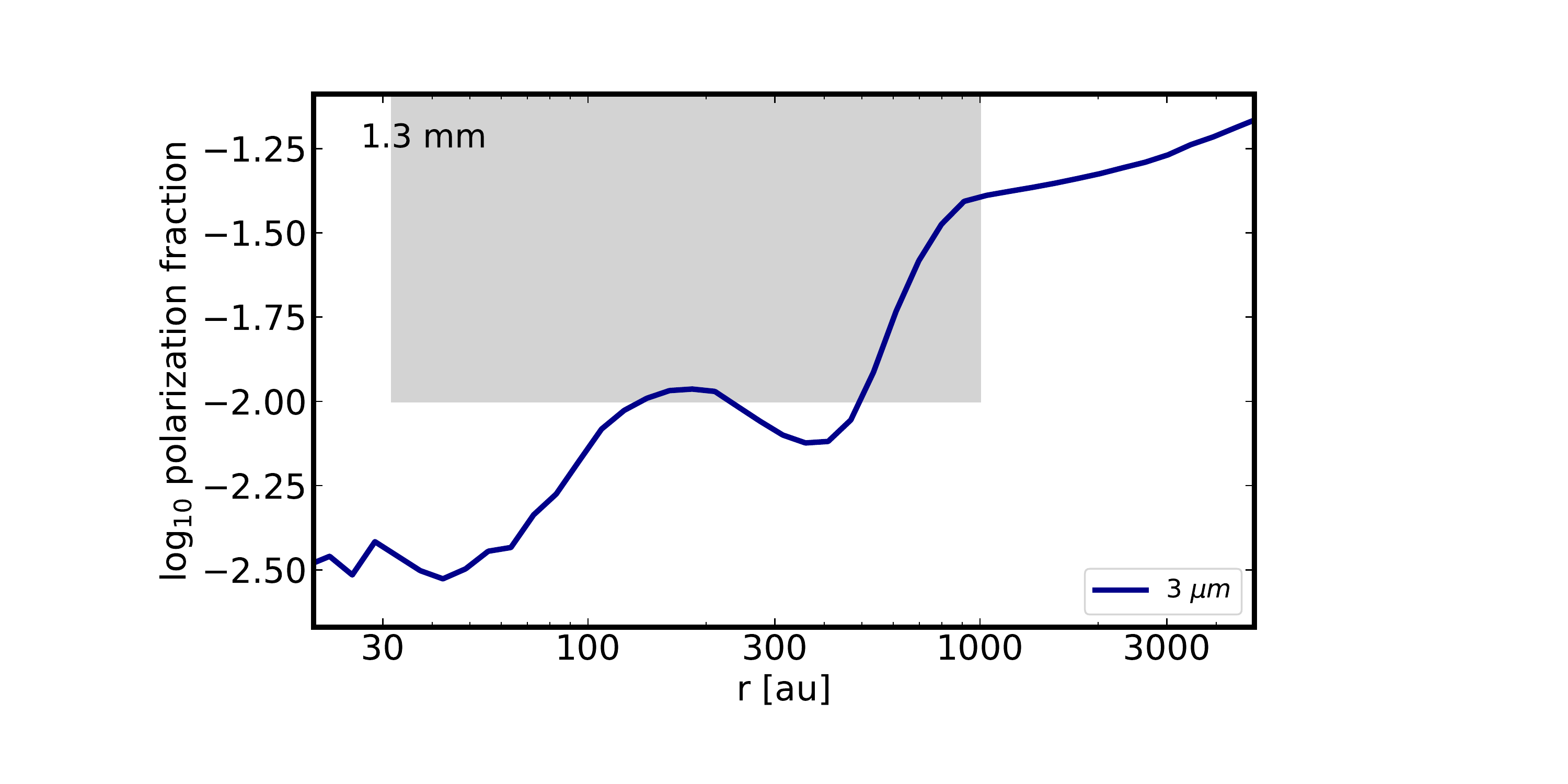}
  \end{tabular}
\caption{Radial mean polarization fraction for the distribution with $a_\mathrm{max} = 3~\micron$, at two wavelengths: $\lambda = 0.8~\mathrm{mm}$ (top panel), and $\lambda = 1.3~\mathrm{mm}$ (bottom panel). In this case the central source is modeled as a blackbody of $10~\mathrm{L_\odot}$ of radius $1~\mathrm{R_\odot}$. The typical values observed in YSOs are shown as a grey region.
This figure shows that even a stronger source with a harder radiation field is not able to significantly increase the amount of polarized dust emission.
}
\label{radial10Lsun}
\end{figure}

\begin{figure}
\centering
\begin{tikzpicture}
\node[above right] (img) at (0,0) {
  % trim left, bottom, right, top
  \includegraphics[width=0.4\textwidth, trim={0.75cm 0.5cm 1.cm 1.2cm},clip]{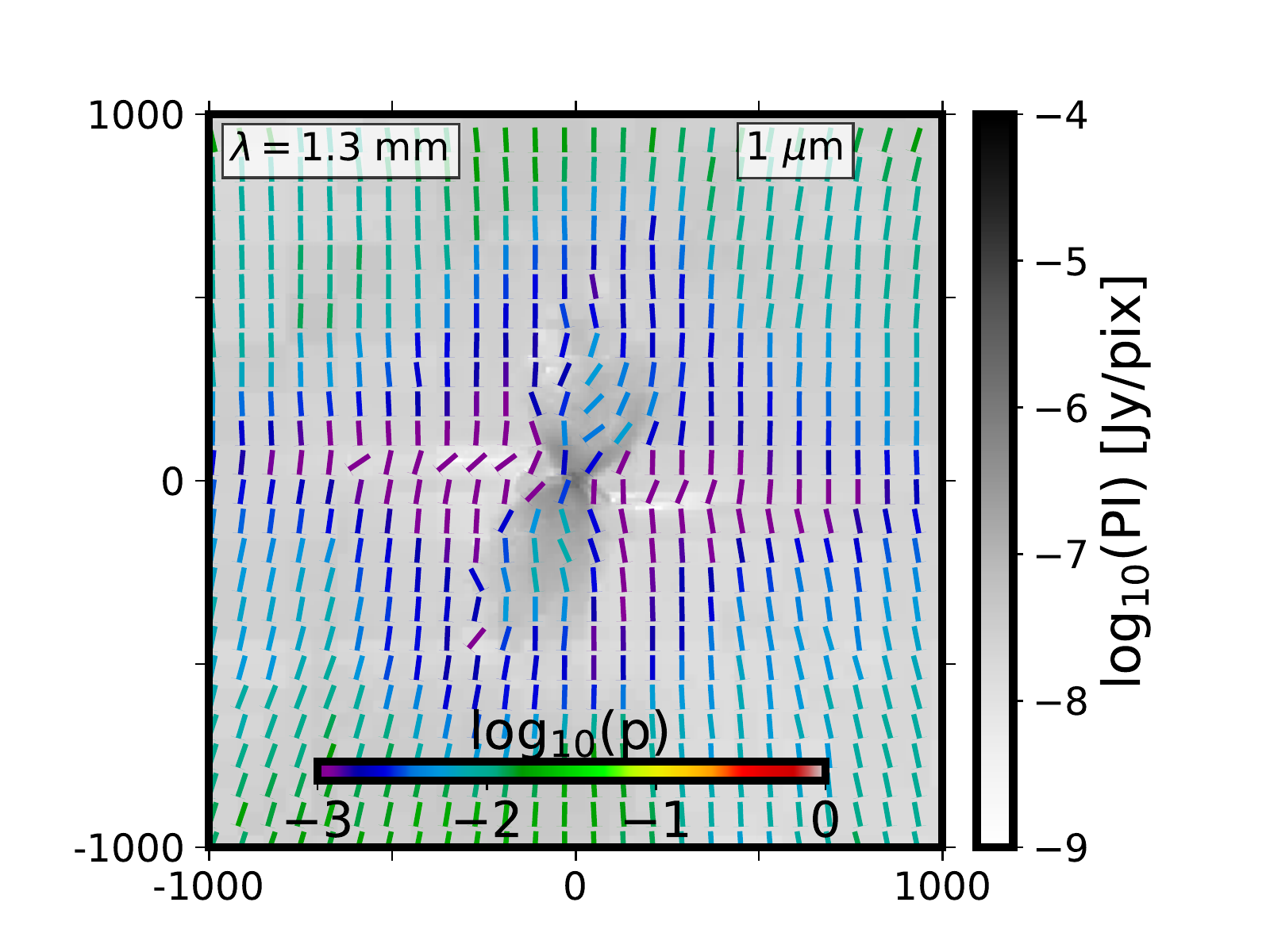}};
  \node [rotate=90, scale=0.8] at (0.3,2.8) {\fontfamily{phv}\selectfont [au]};
    \node [rotate=0, scale=0.8] at (3.35,0.05) {\fontfamily{phv}\selectfont [au]};
\end{tikzpicture}
\caption{Perfect synthetic observation at $\lambda = 1.3~\mathrm{mm}$ for the dust population of small grains ($a_\mathrm{max} = 1~\micron$) composed of only astrosilicates. This figure shows that, not even for this favored dust composition, dust size distributions lacking of dust grains bigger than $10~\micron$ do not reproduce the expected polarization degree.    
}
\label{Silicates}
\end{figure}

%\fi
%%%%%%%%%%%%%%%%%%%%%%%%%%%%%%%%%%%%%%%%%%%%%%%%%%

% Don't change these lines
\bsp	% typesetting comment
\label{lastpage}
\end{document}